\documentclass[fleqn,usenatbib]{mnras}

\usepackage{newtxtext,newtxmath}

\usepackage[T1]{fontenc}
\usepackage{multirow}
\DeclareRobustCommand{\VAN}[3]{#2}
\let\VANthebibliography\thebibliography
\def\thebibliography{\DeclareRobustCommand{\VAN}[3]{##3}\VANthebibliography}

\usepackage{graphicx}	
\usepackage{amsmath}	
\usepackage{xcolor}
\usepackage{listings}
\usepackage[english]{babel}
\usepackage{ulem}
\usepackage{hyperref}
\hypersetup{
    colorlinks=true,
    linkcolor=blue,
    filecolor=magenta,      
    urlcolor=blue,
}
\definecolor{codegreen}{rgb}{0,0.6,0}
\definecolor{codegray}{rgb}{0.5,0.5,0.5}
\definecolor{codepurple}{rgb}{0.58,0,0.82}
\definecolor{backcolour}{rgb}{0.95,0.95,0.92}

\lstdefinestyle{mystyle}{
    backgroundcolor=\color{backcolour},   
    commentstyle=\color{codegreen},
    keywordstyle=\color{magenta},
    numberstyle=\tiny\color{codegray},
    stringstyle=\color{codepurple},
    basicstyle=\ttfamily\footnotesize,
    breakatwhitespace=false,         
    breaklines=true,                 
    captionpos=b,                    
    keepspaces=true,                 
    numbers=left,                    
    numbersep=5pt,                  
    showspaces=false,                
    showstringspaces=false,
    showtabs=false,                  
    tabsize=2
}

\title[The RapidXMM Upper Limit Server]{The RapidXMM Upper Limit Server: X-ray aperture photometry of the \textit{XMM-Newton} archival observations.}

\author[Ruiz et al.]{
A. Ruiz$^{1}$\thanks{E-mail: ruizca@noa.gr}, A. Georgakakis$^{1}$, S. Gerakakis$^2$\thanks{E-mail: gerakakis@planetek.gr},
R. Saxton$^{3}$,
P. Kretschmar$^{4}$,
A. Akylas$^1$,\newauthor\; I. Georgantopoulos$^1$
\\
$^1$Institute for Astronomy \& Astrophysics, National Observatory of Athens, V.  Paulou  \& I.  Metaxa, 11532,  Greece\\
$^2$Planetek Hellas, 44 Kifisias Ave, Athens Space Cluster Bldg C, 15125 Marousi, Athens, Greece\\
$^{3}$TPZ-VEGA for ESA, \textit{XMM-Newton} SOC, ESAC, Apartado 78, 28691, Villanueva de la Cañada, Madrid, Spain\\
$^{4}$ESA-European Space Astronomy Centre, 28691 Villanueva de la Cañada, Madrid, Spain
}

\date{Accepted 2022 January 26. Received 2022 January 26; in original form 2021 June 3}

\pubyear{2022}

\begin{document}
\label{firstpage}
\pagerange{\pageref{firstpage}--\pageref{lastpage}}
\maketitle

\begin{abstract}
This paper presents the construction of the RapidXMM database that is available through the \textit{XMM-Newton} Science Archive and offers access to upper limits and aperture photometry across the field of view of the \textit{XMM-Newton} Pointed and Slew Survey observations. The feature of RapidXMM is speed. It enables the fast retrieval of X-ray upper limits and photometry products in three energy bands  (0.2--2, 2--12, 0.2--12\,keV) for large numbers of input sky positions. This is accomplished using the Hierarchical Equal Area Iso Latitude pixelation of the sphere (HEALPix). The pre-calculated upper-limits and associated X-ray photometry products are reprojected into the HEALPix grid of cells before being ingested into the RapidXMM database. This results in tables of upper limits and aperture photometry within HEALPix cells of size $\approx3$\,arcsec (Pointed Observations) and 6\,arcsec (Slew Survey). The database tables are indexed by the unique integer number of the HEALPix cells. This reduces spatial nearest-neighbour queries by sky position to an integer-matching exercise and significantly accelerates the retrieval of results. We describe in detail the processing steps that lead from the science products available in the \textit{XMM-Newton} archive to a database optimised for sky queries. We also present two simple show-case applications of RapidXMM for scientific studies: searching for variable X-ray sources, and stacking analysis of X-ray faint populations.
\end{abstract}




\section{Introduction}
 
The last two decades have witnessed an explosion in the volume of astronomical observations in the X-ray (0.2--100\,keV) part of the electromagnetic spectrum. This is largely thanks to the smooth and uninterrupted operation of ESA's {\it XMM-Newton}  \citep{Jansen2001} and NASA's {\it Chandra} \citep{Weisskopf2000} X-ray observatories since 1999 as well as the launch of the NuSTAR telescope \citep{Harrison2013} in the early 2010s. Moreover, the continuous scanning of the sky by Gamma-Ray Burst missions such as the Swift-BAT \citep{Gehrels2004} and INTEGRAL \citep{Winkler2003} resulted in the first $4\pi$ surveys at energies $>15$\,keV \citep{Krivonos2010, Oh2018}. The above high-energy missions are supported by archives, which store and serve to the community the accumulating volume of observations in the form of raw data and advanced, science-level products. Nearly all high-energy missions also provide (serendipitous) source catalogues \citep[e.g. 4XMM catalogue;][]{Webb2020}, which list in a table-format the properties of all the sources detected by the telescope/instrument during its operations. They are extensively used by the astronomical community to explore the incidence of X-ray emission in astrophysical populations, search for interesting sources, or carry out statistical studies. 

Despite the undisputed legacy value of source catalogues, there are science applications that extend beyond formal detections. There are astrophysical objects that are hidden in the noise of a given set of observations and yet these non-detections contain important information on their physical properties. The search for variable/transient sources \citep[e.g.][]{DeLuca2021} and their characterisation, e.g. Tidal Disruption Events \citep[e.g.][]{Komossa_Bade1999, Esquej2008, Saxton2019} or Changing Look AGN \citep{Matt2003, LaMassa2015, Ricci2016}, are examples of applications where upper limits, or more broadly sensitivity estimates, play an important role. Such sensitivity calculations are also necessary in statistical investigations of the X-ray properties of sources selected at wavelengths other than X-rays (e.g. optical, near-infrared, radio) to account for the often sizeable X-ray--undetected part of the population. 

The characterisation of non-detected sources, via e.g. the calculation of upper limits, falls under the broad umbrella of X-ray photometry at arbitrary positions within the field of view of a given observation and therefore requires access to imaging data. The HIgh-energy LIght-curve GeneraTor \citep[HILIGT\footnote{\url{http://xmmuls.esac.esa.int/hiligt/}},][]{Saxton2022, Koenig2022} is an example of a webtool that enables such calculations. This, however, is not trivial in the case of large numbers of input source-positions because of the sheer volume of archived observations. We showcase this challenge with the \textit{XMM-Newton} X-ray observatory, which is the focus of this paper. This mission currently provides the largest publicly available X-ray dataset in the energy interval 0.2--12\,keV. The latest release of the \textit{XMM-Newton} serendipitous source catalogue that is based on pointed observations \citep[4XMM-DR9,][]{Webb2020} contains more than half a million unique detections within a footprint that extends over $\rm 1000\,deg^2$. Moreover, the \textit{XMM-Newton} Slew Survey \citep{Saxton2008}, which is carried out as the telescope moves between pointings, currently covers nearly 80\% of the sky and  provides a nearly $4\pi$ X-ray catalogue at the energy range 2--12\,keV. The \textit{XMM-Newton} Science Archive\footnote{\url{http://nxsa.esac.esa.int/nxsa-web}} (XSA) includes both pointed and slew survey products. The pointed observations that use the EPIC cameras \citep[PN, MOS1, MOS2;][]{Turner2001, Struder2001} in imaging (i.e. non-timing) modes are relevant to X-ray photometry. There are a total of 11\,204 such observations taken between 2000 February 3 and 2019 February 26th and included in the 4XMM-DR9 catalogue \citep{Webb2020}, after filtering for data with technical/observational issues (e.g. high particle background). This translates to a rate of about 590 pointed observations per year that are relevant for X-ray photometry applications. Assuming that \textit{XMM-Newton} continues operations for the next 10 years a total of up to 18\,000 pointed observations are expected in the archive by the end of the mission, each with an approximate field of view of $\rm 0.25~deg^2$. In the case of slews, not all of them contain useful data mainly because of high-particle background intervals leading to significant contamination of the whole observation. For example, the 2nd Slew Survey Catalogue includes 2114 slews executed between 2001-08-26 and 2014-12-31. Extrapolating to the end of the \textit{XMM-Newton} mission we may expect up to 5\,000 Slew Observations in the archive, with a typical single slew size of about $30\, \rm deg^2$. 

The above volume of data renders the real-time photometric calculations within the \textit{XMM-Newton} footprint slow. The main overhead is related to input/output time, i.e. accessing the appropriate \textit{XMM-Newton} data products such as images, background and exposure maps in different energy bands. This approach also suffers from redundancy since queries to (nearly) the same position need to be repeated. An obvious solution to overcome these issues is to use look-up tables of pre-calculated photometry that can be used for a broad range of sensitivity calculations, e.g. upper limits. The challenge in this case is to design a storage and query architecture that ensures fast retrievals for positional searches. This is particularly important in the case of many independent positions stored. For the projected number of observations by the end of the \textit{XMM-Newton} mission and a pixel size of 4\,arcsec, there are about $10^{11}$ independent elements (i.e. sky-coordinates) within the Slew Survey footprint (5000 slews of mean size $30\, \rm deg^2$ each) and $4\times10^9$ positions in the case of pointed observations (18\,000 covering an area of $\rm 0.25~deg^2$ each). Nearest position searches over such a large number of records is not trivial and therefore optimisation schemes are essential. 

In this paper we describe the RapidXMM, an analysis pipeline that produces and stores into a carefully designed database both aperture photometry and upper limits for the \textit{XMM-Newton} archival imaging observations. These products are estimated in a dense grid of sky positions with a resolution similar to the pixel size of the \textit{XMM-Newton} images ($\approx 4$\,arcsec). Despite the large number of stored records, the RapidXMM database is optimised to yield fast responses to queries by sky position.

The RapidXMM analysis pipeline has been applied to archival observations and is also used to process new observations as they become available. In Sect.~\ref{sec:db-concept} we motivate the adopted solution for the database architecture and design. The algorithm for the calculation of upper limits is described in Sect.~\ref{sec:algorithm}. The implementation of this algorithm is presented in Sect.~\ref{sec:implemention}, while Sect.~\ref{sec:qa} describes sanity checks to verify the overall quality of the products stored in the database.  A feature of the RapidXMM database is that it enables access to aperture photometry across the footprint of all the observations carried out by  {\it XMM-Newton}. Sect.~\ref{sec:science-applications} demonstrates this potential by presenting show-case science applications of the RapidXMM database. The Appendix \ref{sec:db-fields} presents the database and schema.

\section{Database Concept}
\label{sec:db-concept}

In this work we use the HEALPix pixelisation of the 2-dimensional surface of the sphere \citep{Gorski2005} to accelerate positional searches. The sky is split into equal-area cells using the HEALPix tessellation scheme. Each cell is identified by a unique integer following the {\sc nested}\footnote{\url{https://healpix.sourceforge.io/doc/html/intro.htm}} numbering scheme of HEALPix. There is therefore a one-to-one correspondence between a cell's integer number and the sky coordinate of its centre. This is the feature used in this work to facilitate fast queries at arbitrary sky positions. Upper limits and aperture photometry products are pre-calculated for each pixel of a given {\it XMM-Newton} imaging observation. These are then projected onto the HEALPix grid of cells and stored in a database. This last step associates upper limits within the {\it XMM-Newton} footprint with HEALPix cell integer numbers. Querying the upper limit database by sky position is straightforward. The input sky coordinates are first associated with the integer numbers of the HEALPix cells that contain them. These integers are then matched against the HEALPix cell numbers stored in the database to return any upper limits associated with them. The sky-coordinate nearest-neighbour search is therefore reduced to an integer matching exercise. An important parameter in this scheme is the size of the HEALPix cell that controls the spatial resolution of the database. In HEALPix terminology the relevant parameters are the {\sc order} or {\sc nside} of the sky tessellation. The values of these parameters determine the number of independent upper limit estimates that will be stored in the database. For example a HEALPix {\sc order} of 16 ({\sc nside}=$2^{16}$) corresponds to a resolution of about 3\,arcsec, similar to the pixel size of \textit{XMM-Newton}. The decision on the HEALPix resolution is a trade-off between accurately describing spatial variations across the sky and size of the database. 

\begin{figure*}
\begin{center}
\includegraphics[angle=0, width=\textwidth]{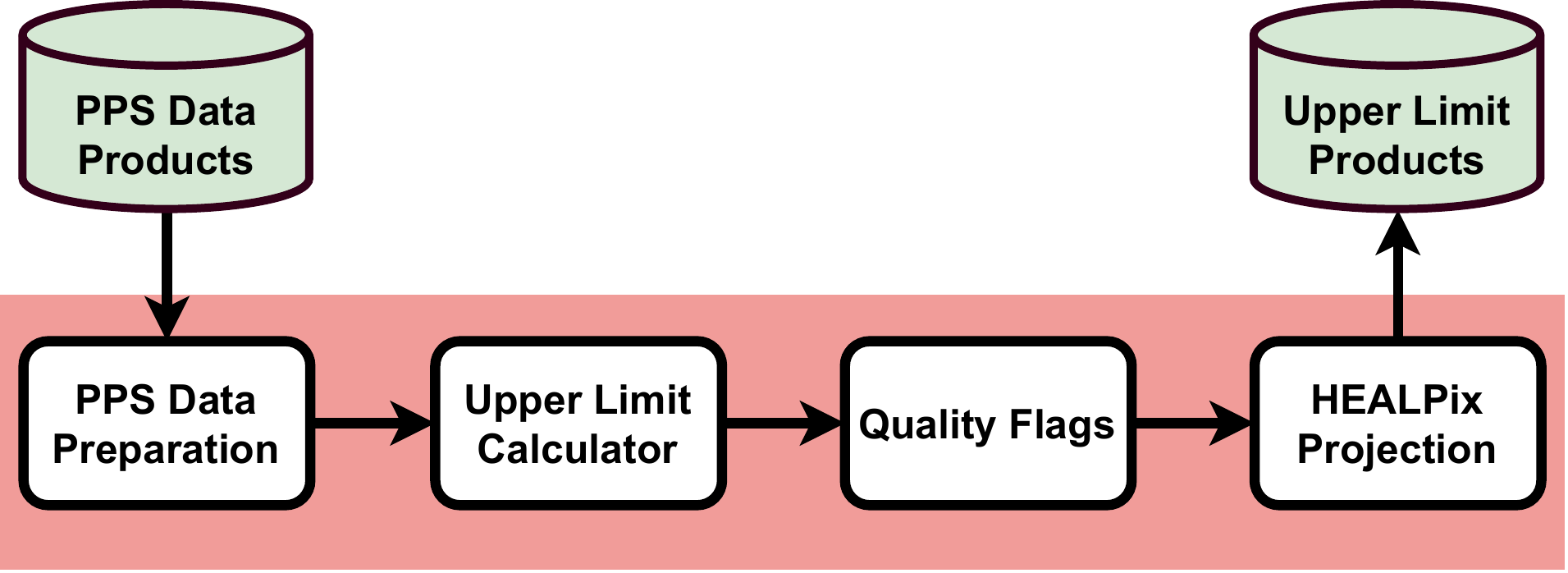}
\end{center}
\caption{Flow chart presenting the RapidXMM data analysis process that estimates upper limits and associated products. Each rectangle within the pink shaded region corresponds to a calculation module. The green cylinders signify data fed into the analysis flow (Processing Pipeline Subsystem products) or returned from it (Upper Limit Products). The first step of the analysis flow chart is the preparation of the PPS (Processing Pipeline Subsystem) data (see Sect.~\ref{sec:data-curation}). This is followed by the Upper Limit calculation module (Sect.~\ref{sec:calc}), the quality flag estimation (Sect.~\ref{sec:quality-flag}) and the reprojection onto the HEALPix grid (Sect.~\ref{sec:healpix-repro}).}\label{fig:analysis-flow}
\end{figure*}

\section{The Upper-limit estimation algorithm}
\label{sec:algorithm}

 The estimation of upper limits is based on aperture photometry. The advantage of this approach is that the observed number of photons within the aperture follows the well-known Poisson probability distribution with an expectation value that depends on both the background level and any underlying source contribution. The algorithm for converting observed number of photons to sources' upper limits at a given confidence interval is based on the work of \citet{Kraft1991} and is described below. 

\begin{figure*}
\begin{center}
\includegraphics[angle=0, width=\textwidth]{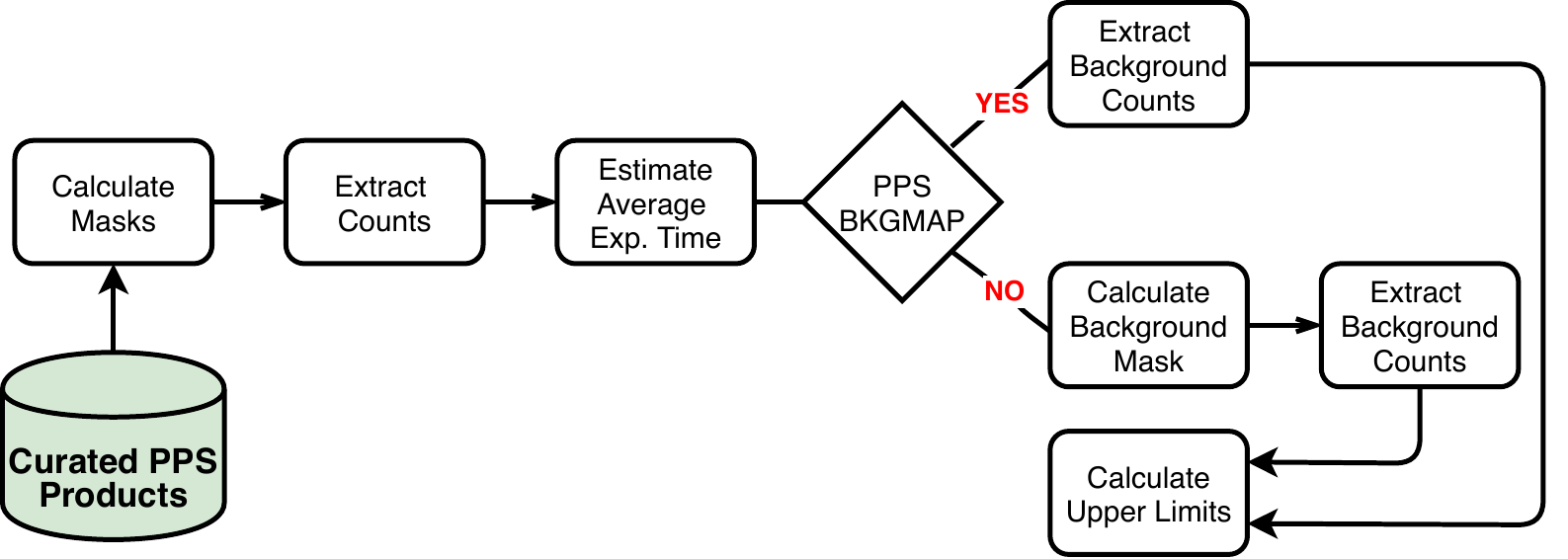}
\end{center}
\caption{Flow chart of the Upper Limit module. The rectangles show different analysis steps. The diamond-shaped boxes correspond to decision points in the process. The  green cylinder signifies data fed into the analysis flow. The module starts with the calculation of masks (observation, source masks; see Sect.~\ref{sec:calc}), proceeds with the aperture count extraction and the calculation of the mean exposure time within the extraction kernel. This is then followed with the determination of the background level within the aperture. At this stage the flow forks into two branches. If PPS background maps are available (Pointed Observations) then these are used to determine the background level. Otherwise the background is determined from the science image. This requires masking non-exposed pixels and regions contaminated by nearby sources (background mask). The background level is then determined from an annulus centred at the position of interest. Finally the products determined in the steps above (extracted counts, exposure time, background level) are combined to determine upper limits (see equations of Sect.~\ref{sec:algorithm}).}
\label{fig:UL-flow}
\end{figure*}

Suppose an X-ray observation with a given exposure time and detector characteristics, e.g. Point Spread Function, background level. Given a position within the observation footprint we need to estimate the count-rate upper-limit a source could have had. We choose to approach this problem using aperture photometry. Suppose an aperture of radius $R$ centred at the position of interest. The total number of photons within this aperture, $N$, consists of a background component, $B$, and a possible source contribution, $S$. The level of the background expectation value, $B$, within the aperture can be estimated from source-free regions close to the position of interest. The source contribution to the observed number of photons in the aperture depends on the shape of the Point Spread Function (PSF) and the exposure time of the observation. If the putative source has a count rate CR then
\begin{equation}
S = \mathrm{CR} \cdot \sum_{i,j} t_{i,j} \cdot P_{i,j},
\end{equation}
where $t_{i,j}$ are the values of the (vignetted) exposure map and $P_{i,j}$ the values of the normalised PSF at the pixel $(i,j)$. The summation is over all pixels within the extraction radius. Assuming that the exposure map varies slowly within the extraction area\footnote{This is a reasonable assumption for relatively small values of $R$, except at positions near the detector gaps and field edges. The latter effects are taken into account using pixel masks as explained in Sect.~\ref{sec:calc}} then
\begin{equation}
S \approx \mathrm{CR} \cdot t \cdot \sum_{i,j} P_{i,j} = \mathrm{CR} \cdot t \cdot EEF,
\end{equation}
where EEF is the Encircled Energy Fraction and measures the fraction of total source photons within the aperture $R$. The EEF depends on the shape and extent of the X-ray PSF and typically varies with position on the detector and photon energy. 
The observed number of photons $N$ in the aperture is a Poisson realisation with expectation value $T = S + B$
\begin{equation}\label{eq:poisson}
P(N|S+B) = \frac{(S+B)^N\,e^{S+B}}{N!},
\end{equation}
where $P(N|S+B)$ is the probability of observing $N$ photons given the expectation value $(S+B)$. Next we assume that $B$ can be determined to a very high level of accuracy. This is a reasonable assumption in the case of the background-limited observations with the EPIC cameras of \textit{XMM-Newton}. Source confusion is typically low in these observations and therefore sufficiently large source-free detector regions are available to provide reliable estimates of the local background level. Under this assumption it is possible to integrate the probability density function of Eq.~\ref{eq:poisson} with respect to $S$ and determine the corresponding cumulative distribution as a function of this parameter.  The upper limit, $UL$, at an arbitrary confidence interval CL is then defined as the value of $S$ for which the cumulative probability equals CL. Following the seminal work of \citet[][]{Kraft1991} this can be expressed as
\begin{equation}\label{eq:confidence}
C \cdot \int_{0}^{ \mathrm{UL}}\frac{(S+B)^N\,e^{S+B}}{N!}\, dS =  \mathrm{CL},     
\end{equation}
where $C$ is the constant of normalisation. Assuming a prior where the only condition is for the UL to be positive, $C$ is defined as
\begin{equation}\label{eq:norm}
C^{-1}
= \int_{0}^{\infty}\frac{(S+B)^N\,e^{S+B}}{N!}\, dS
= \int_{B}^{\infty}\frac{T^N\,e^T}{N!}\, dT
= \Gamma(N+1, B).
\end{equation}
In the equation above the integration has been expressed in terms of $T=S+B$ and $\Gamma(a, x)$ is the upper incomplete gamma function. We can also change the integration variable to  $T=S+B$ in Equation \ref{eq:confidence} and replace the normalisation constant from Equation \ref{eq:norm}. Equation \ref{eq:confidence} can then be rewritten
\begin{equation*}
\int_{B}^{B+ \mathrm{UL}}\frac{T^N\,e^T}{N!}\, dT = \mathrm{CL}\cdot C^{-1} \Rightarrow
\end{equation*}
\begin{equation*}
\int_{0}^{B+ \mathrm{UL}}\frac{T^N\,e^T}{N!}\, dT -  \int_{0}^{B}\frac{T^N\,e^T}{N!}\, dT = \mathrm{CL}\cdot C^{-1} \Rightarrow
\end{equation*}
\begin{equation}
\begin{split}
\gamma(N+1, \mathrm{UL}+B) - \gamma(N+1, B) = \mathrm{CL}\cdot \Gamma(N+1, B) & \Rightarrow \\
\gamma(N+1, \mathrm{UL}+B) = \gamma(N+1, B) + \mathrm{CL}\cdot \Gamma(N+1, B), & \\
\end{split}
\end{equation}
where $\gamma(a, x)$ is the lower incomplete gamma function.

\begin{table*}
\centering
\caption{Parameters used in the calculation of upper limits for Pointed and Slew Survey observations. The columns are: (1) the observation type, Pointed or Slew, (2) the radius $R$ in units of arcseconds within which counts are extracted from the science images, (3) the inner/outer radius of the annulus in units of arcseconds used to extract background counts (4) the Encircled Energy Fractions (EEFs) for each radius $R$. The EEF is different for each EPIC detector. The Slew Survey uses the EPIC-PN only.}
\begin{tabular}{ c c c  ccc}
\hline 
Observation & Source & Background  annulus &  \multicolumn{3}{c}{Encircled Energy}\\
 type       & extraction radius            & inner/outer radius   & \multicolumn{3}{c}{Fraction (EEF)}        \\
            & (arcsec)          &  (arcsec)            &    PN & MOS1 & MOS2        \\
\hline
Pointed     & 15  & 0/60  & 0.68 &	0.69 & 0.71 \\
Slew        & 30  & 60/180 & 0.85 & -- & -- \\ 
\hline
\end{tabular}
\label{tab:params}
\end{table*}

The upper limit at the confidence interval CL can be estimated  by inverting the lower gamma function on the left hand side:
\begin{equation}\label{eq:ULgamma}
 \mathrm{UL} = \gamma^{-1}(N+1,  \mathrm{CL} \cdot \Gamma(N+1, B) + \gamma(N+1, B)) - B.
\end{equation}
The UL is in units of counts. It can be converted to count-rate by dividing with the exposure time $t$ and  the EEF to correct for the fraction of source photons outside the aperture $R$ 
\begin{equation}\label{eq:UL}
 \mathrm{CR}_{ \mathrm{UL}} =  \mathrm{UL} / (t \cdot \mathrm{EEF}). 
\end{equation}
The relation above corresponds to the count-rate upper limit in the case of a single detector, i.e. one of the EPIC PN, MOS1 or MOS2. This equation is used by the RapidXMM to estimate upper limits. It is some times desirable however, to combine information from different cameras to obtain a deeper and possibly more physically interesting upper limit. The issue that needs to be resolved in this case is the different energy response of the EPIC cameras. This effect can only be accounted for in the context of a spectral model to yield an upper-limit of a source in X-ray flux rather than count rate. In this case some of the equations above need to be modified. The total number of counts within the extraction aperture from all three detectors is
\begin{equation}\label{eq:total_all_cameras} 
N = f_X \cdot  \sum_{i=1}^{3} t_i\cdot  \mathrm{ECF}_i \cdot   \mathrm{EEF}_i + \sum_{i=1}^{3} B_i, 
\end{equation}
where $f_X$ is the source's flux in a given energy band, $t_i$,  $B_i$,  $\mathrm{EEF}_i$ are the exposure time, background level and encircled energy fraction of the camera $i$, and  $\mathrm{ECF}_i$ is the energy to count conversion factor of the detector $i$.  The latter depends on the adopted spectral model of the source and the energy response of the instrument. The left-hand side of Equation \ref{eq:ULgamma} should also be modified by substituting the UL term (units of counts) from Equation \ref{eq:total_all_cameras}. As a result the flux upper limit is 
\begin{equation}\label{eq:flux_all_cameras} 
f_X =  \frac{\mathrm{UL}}{\sum_{i=1}^{3} t_i \cdot   \mathrm{ECF}_i \cdot   \mathrm{EEF}_i},
\end{equation}
and UL is given by Equation \ref{eq:ULgamma} after substituting $B$ with $\sum B_i$. The RapidXMM database does not store upper limits for the combined EPIC cameras but provides all the necessary products (counts, exposure time, background level), which when combined with a spectral model (i.e. ECFs) can yield flux upper limits via Equation \ref{eq:flux_all_cameras}.

\section{The RapidXMM Implementation}
\label{sec:implemention}

In Sect.~\ref{sec:algorithm} we show how count-rate upper limits can be estimated at a given position within an X-ray image from known quantities ($N$, $B$, $\mathrm{CL}$, $t$, and $ \mathrm{EFF}$) in terms of incomplete gamma functions. There are a number of implementations of these functions in different programming languages that allow fast calculations. The RapidXMM project is based on Python \citep{VanRossum09} and uses the fast, vectorized implementation of gamma functions included in the SciPy module \citep{2020SciPy}.  

A flow chart of the RapidXMM data processing algorithm for a given \textit{XMM-Newton} observation is shown in Fig.~\ref{fig:analysis-flow}. The science observations are downloaded from the XSA archive and, if needed, additional products are calculated for the RapidXMM energy bands (Sect.~\ref{sec:data-curation}). These products (images, background maps and exposure maps) are used to extract the information needed to estimate upper limits via Eqs.~\ref{eq:ULgamma} and~\ref{eq:UL} (Sect.~\ref{sec:calc}). Quality flags are assigned to each upper limit according to the number of counts and the fraction of good (exposed) pixels contained in the corresponding extraction regions (Sect.~\ref{sec:quality-flag}). Finally, the upper limits calculated for each image pixel, and their ancillary photometric information (extracted counts, background level, mean exposure time), are reprojected into a HEALPix grid covering the image (Sect.~\ref{sec:healpix-repro}) and ingested into the RapidXMM database (Sect.~\ref{sec:rapidxmm-db}). Each of these steps is described in detail in the following sections. 

\begin{figure*}
\begin{center}
\includegraphics[width=\linewidth]{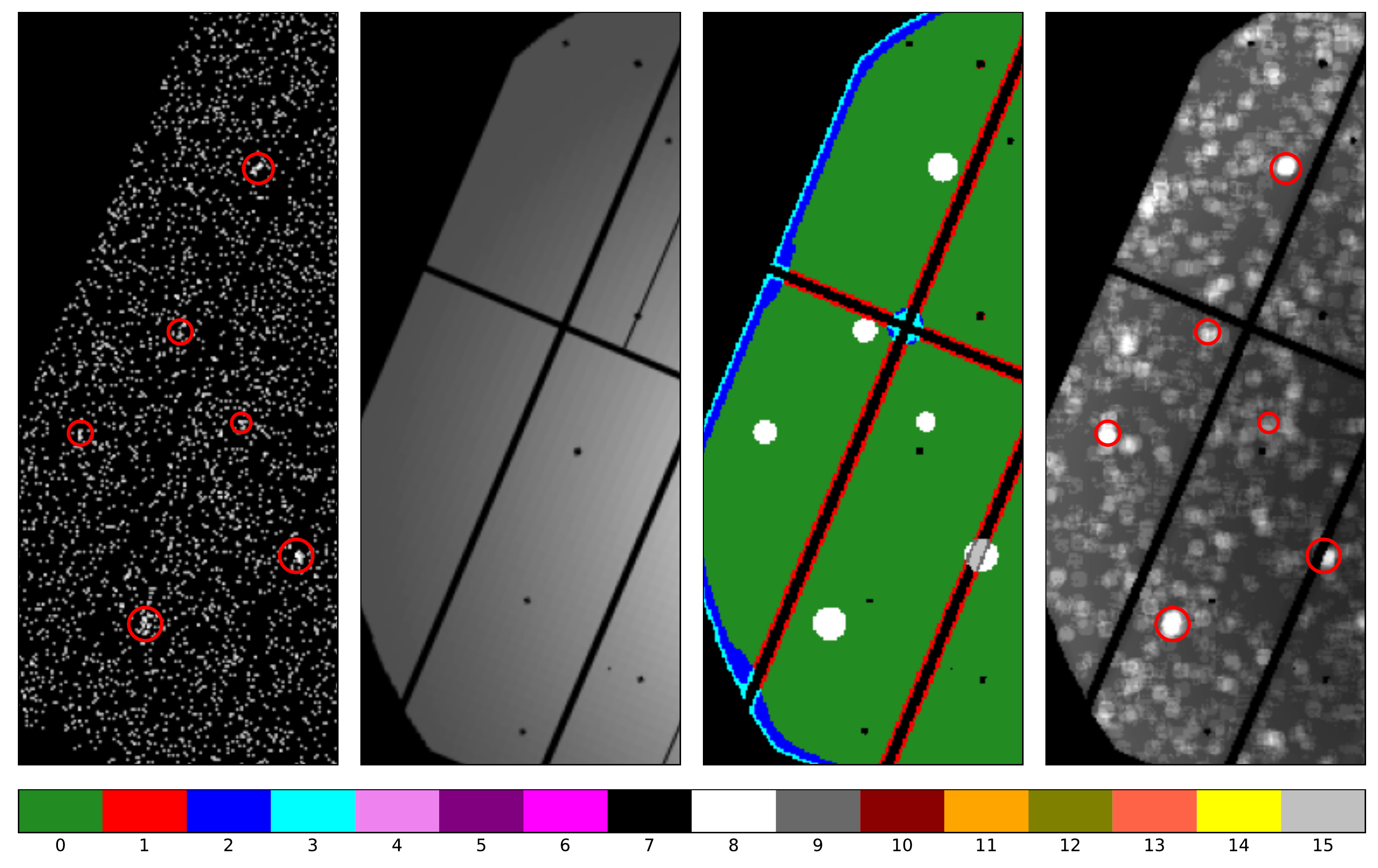}
\end{center}
\caption{Demonstration of the quality flag products generated by the RapidXMM. The left panel is the 0.2--2~keV EPIC-PN image of the Pointed Observation with Obs.ID number 0404967501. Only a section of the image is shown to highlight details. The red circles mark the positions of sources detected in this observation by the PPS. The radius of the each circle depends on the apparent brightness of each source. The second panel from the left shows the exposure map of this observation. Darker regions correspond to lower exposure time and black identifies non-exposed areas, e.g. outside the field of view, CCD gaps or bad pixels. The third panel from the left is the RapidXMM Quality Flag image of this observation. Each pixel on that image has a decimal integer value between 0 and 15 that depends on the flag bits raised during the estimation of upper limits (see Table \ref{tab:flags}). The correspondence between flag value and colour is shown by the colour-map at the bottom. Green corresponds to non-flagged pixels (flag value 0). White corresponds to a flag value of 8 and marks the positions of sources. The boundaries of the image are also flagged (values between 1 and 3 depending on the bits raised). The panel on the right shows the image of the $2\sigma$ count-rate upper limits in the 0.2--2~keV band. X-ray detected sources are highlighted with red colour. Dark colour corresponds to lower count-rate values. The observed gradient from the middle to the edges is the result of the vignetting. The apparent granularity of the upper limits across the field of view is because of the Poisson nature of the \textit{XMM-Newton} background, i.e. the fact that the number of photons included within the extraction aperture takes only integer values.}\label{fig:flag}
\end{figure*}

\subsection{Data preparation}
\label{sec:data-curation}
The analysis of a particular \textit{XMM-Newton} observation (Pointed or Slew) starts by downloading the relevant data from the XSA. The RapidXMM analysis is based on the science products generated by the \textit{XMM-Newton} Pipeline Processing Subsystem (PPS). The calculation of upper limits is only relevant to the imaging observations carried out by the \textit{XMM-Newton}, i.e. those obtained by the European Photon Imaging Cameras \citep[EPIC,][]{Struder2001, Turner2001}. The relevant PPS products include images, background maps and exposure maps in the various PPS energy bands for the three EPIC instruments (PN, MOS1, MOS2), as well as the corresponding X-ray source catalogue. The Slew Survey observations are using the EPIC-PN instrument only. Also, the Slew PPS products do not include background maps. In this case the local background value at a given position within the observation footprint is estimated  on the fly from the RapidXMM code. 

In the case of \textit{XMM-Newton} Pointed Observations the PPS products need to be further analysed before calculating upper limits. This is because the energy bands used by RapidXMM and those adopted by the PPS are different. RapidXMM estimates upper limits in three broad energy bands, 0.2--2 (soft), 2--12 (hard) and 0.2--12\,keV (full). With the exception of the full-band, the PPS generates Pointed Observation products in  narrow energy intervals \citep{Webb2020}.\footnote{\url{http://xmmssc.irap.omp.eu/Catalogue/4XMM-DR9/4XMM-DR9\_Catalogue\_User\_Guide.html\#TabBands}} Therefore, the Pointed Observation PPS products have to be merged to produce images in the RapidXMM broad bands. The RapidXMM soft-band (0.2--2\,keV) is constructed from the PPS bands 1 (0.2--0.5keV), 2 (0.5--1\,keV) and 3 (1--2\,keV). The RapidXMM hard-band (2--12\,keV) is synthesised from the PPS bands 4 (2--4.5\,keV) and 5 (4.5--12\,keV). The RapidXMM full-band is build from the PPS bands 1 to 5.\footnote{PPS products include images for the energy band 8 that corresponds to the RapidXMM full-band. However the PPS does not provide background maps in that energy band. For consistency, we decided to calculate the RapidXMM full-band images from the narrow band PPS images.} The RapidXMM data preparation  module merges the Pointed Observation PPS narrow bands to generate the RapidXMM soft and hard bands. In the case of images and background maps this  operation is the addition of the individual components. For exposure maps the narrow-band components are averaged. For Slew Observations the PPS products are generated in the same broad energy bands used by RapidXMM, i.e. 0.2--2, 2--12 and 0.2--12\,keV. In this case there is no need to further process the PPS data. 

Each distinct \textit{XMM-Newton} observation is characterised by a unique identification number (Obs.ID). Pointed observations include at least one exposure for each of the \textit{XMM-Newton} instruments. Slew Observations typically have long footprints that extend over many degrees on the sky. For practical reasons the PPS splits them into multiple patches. Each Obs.ID is analysed separately by the RapidXMM. Upper limits are determined independently for each exposure or for each patch of a given Obs.ID.

\subsection{Upper Limit Calculation}
\label{sec:calc}
The estimation of upper limits via Eq.~\ref{eq:ULgamma} requires knowledge of the total extracted photons within an aperture, the expected background contribution to the observed counts and the exposure time at that position. The calculation module of RapidXMM determines these quantities for each pixel position of individual exposures or patches within an Obs.ID and for each of the three RapidXMM energy bands. Upper limits are estimated at three confidence intervals that correspond to probabilities of 84.13, 97.72 and 99.87\% that the true count-rate value lies below the respective upper limit. The probabilities above correspond to the one-sided $1\sigma$, $2\sigma$ and $3\sigma$ confidence levels of a Normal distribution. It is recognised that the use of the one-sided probability definitions above is not common. The calculation of upper limits however, involves an integration that is essentially bounded only on one side (Eq.~\ref{eq:confidence}). It therefore makes sense in this case to express the confidence interval in only one direction.

At a given pixel position, image counts are integrated within a circular aperture of radius $R$. The expected background level within that aperture is estimated from either the PPS background map (Pointed Observations) or the image itself (Slew Survey). Background counts are extracted in an annulus of inner and outer radii $R_{in}$,  $R_{out}$ respectively. The adopted values of $R$, $R_{in}$ and  $R_{out}$ are shown in Table \ref{tab:params}. The average exposure time within the aperture $R$ is also estimated using the corresponding exposure map. These calculations account for non-exposed pixels within the extraction apertures, e.g. CCD gaps (Pointed observations only), observation field of view, hot/bad pixels (see below for details). If the background is determined from the science image rather than the PPS background maps (e.g. Slew observations), areas in the vicinity of detected sources are masked out to avoid contamination. All the count extraction operations are equivalent to the convolution of the input image with a kernel that matches the aperture/annulus size. RapidXMM uses Fast Fourier Transforms to perform convolutions and significantly accelerate the calculations. The prescription outlined above is described in more detail below and is graphically demonstrated in Fig.~\ref{fig:UL-flow}. 

\begin{table*}
\centering
\caption{
Upper limit quality flag bits.  The first column lists the bit number and the second  column is the corresponding flag value. The third column provides a short description of the flag-raising condition.}
\begin{tabular}{ c c l}
\hline 
Bit Number & Flag Value & Description  \\
\hline
0   & 1 & Fraction of exposed pixels within the count extraction aperture is lower than 0.75\\
1   & 2 & Fraction of good pixels within the background extraction aperture is lower than 0.7 (Pointed) or 0.5 (Slew) \\
2   & 4 & Total counts in the background extraction region is zero  (Pointed only)\\
3   & 8 & Pixels lies in the vicinity of a source \\
\hline
\end{tabular}
\label{tab:flags}
\end{table*}

First RapidXMM calculates two sets of masks, one that describes the exposed area of a given observation (observation masks) and a second one that masks regions within which the observed counts are dominated by X-ray point-source detections (source masks). The exposure map is used to identify pixels with positive exposure time and generate "Observation" masks. These are 2-dimensional integer images with pixels that take value of either one in exposed areas or zero otherwise. The latter case corresponds to pixels in CCD gaps, outside \textit{XMM-Newton}'s Field-of-View (FoV), or known to be hot/bad. There is one Observation mask for each exposure or patch of an Obs.ID. These masks are used to estimate the fraction of non-exposed pixels in an aperture at the count-extraction stage of the calculations. They are also  employed to flag upper-limits on pixels that lie close to CCD edges. "Source" masks are integer images with pixel values of unity in the vicinity of X-ray detected sources and zero everywhere else. Areas close to X-ray detections are identified using the source positions and source radii provided by the PPS products in the form of DS9 regions files. These files are created by the \texttt{slconv} SAS task,\footnote{\url{http://xmm-tools.cosmos.esa.int/external/sas/current/doc/slconv/slconv.html}} using a circular shape at the positions of the sources detected by \texttt{emldetect} in the observation, and with radius determined from the sum of the source count rates for each narrow energy band (1--5) and detector. The source radius describes the extent of the area within which the observed photons are dominated by that source. The Slew observation PPS products do not include information on the radii of any detected sources. In this case a fixed radius of 45\,arcsec is used to define source-contaminated areas. Source masks are used to flag upper-limits in the vicinity of X-ray detections, for which contamination from nearby sources may be an issue. The observation and source masks are further combined to define "background" masks. These are defined to be the observation masks with additional pixels set to zero in source-dominated regions.  The background masks are used to exclude potentially contaminating source photons from the calculation of the background expectation value at a given position in the case of the Slew Survey observations, for which no PPS background maps are provided. 

The mask construction is followed by the count-extraction stage. The science images of a given Obs.ID at a given energy band are convolved with a circular kernel of radius $R$ (15\,arcsec for pointed observations, 30\,arcsec for Slews; see Table~\ref{tab:params}). Pixels with zero value on the observation mask are also set to zero (i.e. the science image is multiplied by the observation mask) and hence, are effectively ignored by the convolution operation. The convolution of the same kernel with the observation mask yields an estimate of the area in pixels, $A_{R}$, of the extraction aperture excluding non-exposed pixels. The same convolution operation is also applied to the exposure maps. In these case however, the integrated exposure time within the aperture (i.e. the result of the convolution) is divided by the area $A_R$ of the extraction kernel to calculate averages.  

The next step is the estimation of the background level within the aperture with area $A_R$. Two different approaches are followed in the case of Pointed and Slew observations. For the former the PPS-generated background maps are used to measure the expected number of background counts within the aperture of radius $R$. The PPS background maps are convolved with a circular kernel of 60\,arcsec radius.  The convolution of the observation mask with the same kernel yields the kernel’s area in pixels, $A_B$, at each position corrected for non-exposed pixels. Scaling this convolved image by the area ratio $A_R$/$A_B$, yields an estimate of the background expectation in the aperture of radius $R$ and area $A_R$. The PPS products of Slew observations do not include background maps. In this case the background level is determined from the science images after multiplying them with the source mask to effectively nullify pixels in the vicinity of detected sources, CCD gaps etc. The same approach as above is adopted with the exception that the convolution kernel is an annulus with inner and outer radii of 60 and 180\,arcsec respectively. The area of the kernel is measured using the background mask and this is then used to scale the extracted counts to the area of the aperture with radius $R$.

The result of count-extraction stage are images of integrated counts, background count expectations and mean exposure times within an aperture of radius $R$. These are fed to  Eq.~\ref{eq:ULgamma} to determine images of upper limit count rates for the each of the confidence level probabilities 84.13\% ($1\sigma$), 97.72\% ($2\sigma$) and 99.87\% ($3\sigma$). Table \ref{tab:params} presents the adopted Encircled Energy Fraction (EEF) for each EPIC camera and observation type (pointed or slew). These EEFs are the same as those used in the {\sc eupper} task of \textit{XMM-Newton}'s SAS (Science Analysis System) and correspond to the Point Spread Function (PSF) size at 1.5\,keV. Ideally energy and off-axis angle dependent EEFs should be used. The \textit{XMM-Newton} EPIC cameras however, have been designed to minimise the variation of the PSF, and hence the EEF, across the field of view and across the range of observed photon energies. The adopted EEFs therefore provide an adequate (within 5\%, see {\sc eupper} documentation) representation of the fraction of source photons inside the extraction radius.

\subsection{Quality flags}
\label{sec:quality-flag}
RapidXMM estimates quality flags that characterise the upper limit at a given pixel and identify potential issues, e.g. proximity of the pixel in question to the edge of the field-of-view or X-ray sources. These flags are stored as bit values. There are a total of four flags which are defined in Table \ref{tab:flags} and described in detail below:
\begin{itemize}
    \item The first bit (number 0) is raised if the fraction of good pixels (i.e. positive exposure time based on the observation mask) within the count-extraction aperture falls below 75\%. This may occur for example, close to CCD edges or in regions with bad pixels. The choice of threshold is empirically determined to allow this flag to trace in a satisfactory manner CCD edges (see Fig.~\ref{fig:flag}).
    
    \item The second bit (number 1) is raised if the fraction of good pixels within the background-extraction region falls below a threshold. For Pointed observations the observation masks is used to define good pixels (positive exposure time). For this type of observation the threshold below which the flags is raised is set to 70\%. In the case of Slew observations the background is estimated from the image itself after masking out detections, as explained above, and the background extraction area is much larger than that of Pointed observations. As a result in this case the background mask is used to define good pixels (positive exposure and not in the vicinity of detections). The threshold above which the flag is raised is also different from that of the Pointed observations and is set to 50\%. The threshold values are determined empirically so that they trace the edge of the Field-Of-View of a given observation (see Fig.~\ref{fig:flag}).
    
    \item The third bit (number 2) is raised if the background region contains zero counts. This occurs often in the soft band (0.2--2keV) of the Slew Survey observations, i.e.about 40--50\% of the pixels in these data fall under this category. It should be emphasised that a zero value for the background is not a problem for the determination of upper limits. For that reason it was decided to switch this flag off and not raise it in the case of Slew Survey observations.  This flag remains nevertheless useful in the case of Pointed observations, since a zero background in this case may indicate potential issues. 
    
    \item The fourth bit (number 3) is raised if the pixel lies close to a detected source. The proximity of a pixel to an X-ray detection is determined using the source mask. If for a given pixel there is overlap between the source extraction aperture of radius $R$ and the source mask then the flag is raised. Some XMM Slew observations do not contain a PPS source list. No source flagging is done in this case.  
\end{itemize}

Before passing the generated data to the next stage of the RapidXMM pipeline two sanity checks are carried out. It is first tested if any of the upper limits associated with exposed pixels on an image has a value of zero, infinity or NaN (Not-a-Number). Such values should not occur and indicate potentially problematic observations or non-anticipated problems during the calculation. The second test checks if all the upper limits on image have a flag raised. This test catches instances of problematic observations in the context of upper limit calculation. This may happen for example, in the case of small window-mode observations, in which only the central CCD window is exposed. The available area in this case is small and it may happen that all pixels are flagged because of insufficient source-free regions for the background estimation. If any of these tests is positive, the process is interrupted and the data corresponding to this image are rejected. We estimate that these issues affect about 0.4\% of all observations. Additionally there are Obs.IDs, for which the PPS, for various technical reasons, cannot produce images, exposure maps and/or background maps (relevant to Pointed observations only) for some or all the exposures within the observation. If any of these products is missing no upper limits can be calculated for the relevant exposure.  

\begin{figure}
\begin{center}
\includegraphics[angle=0,height=.6\columnwidth]{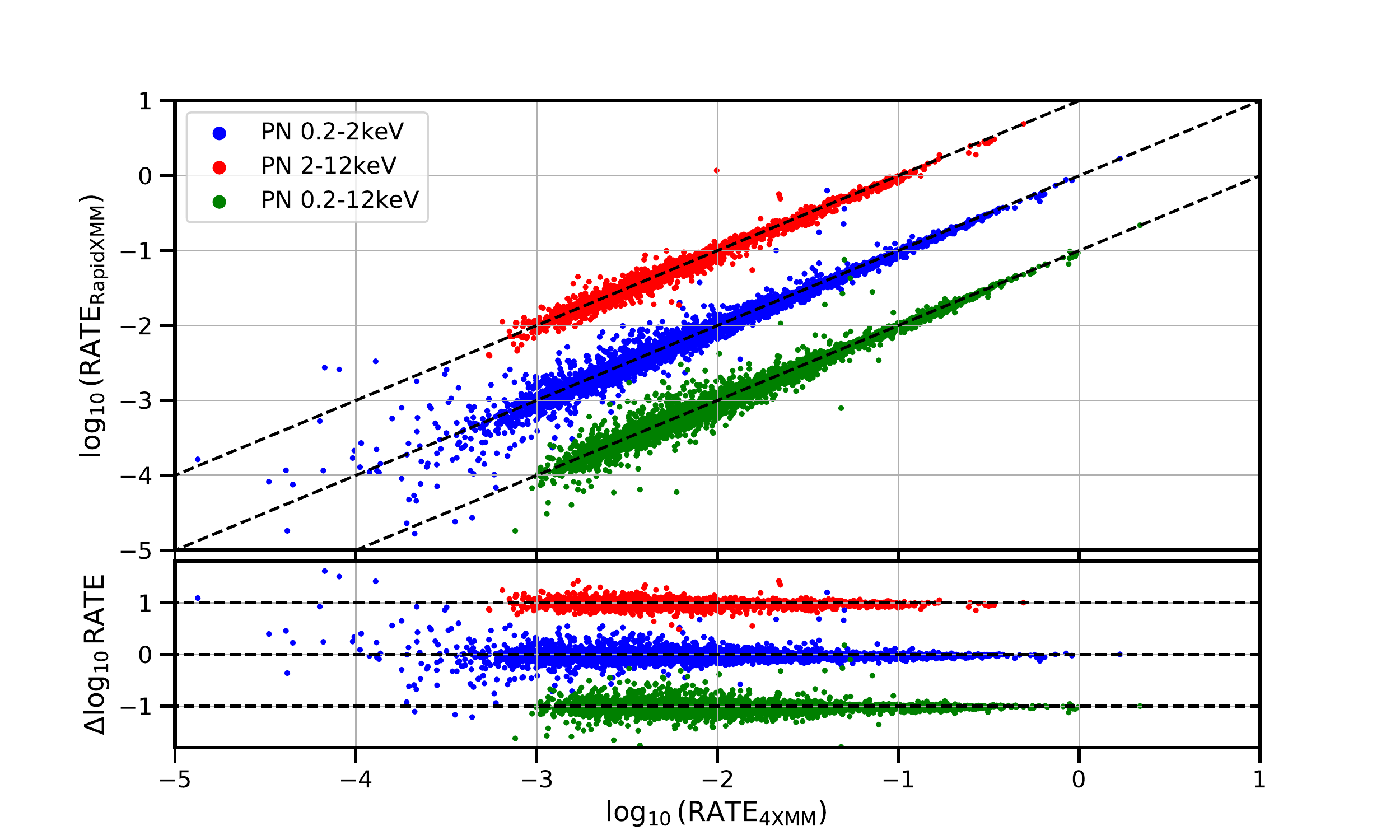}
\includegraphics[angle=0,height=.6\columnwidth]{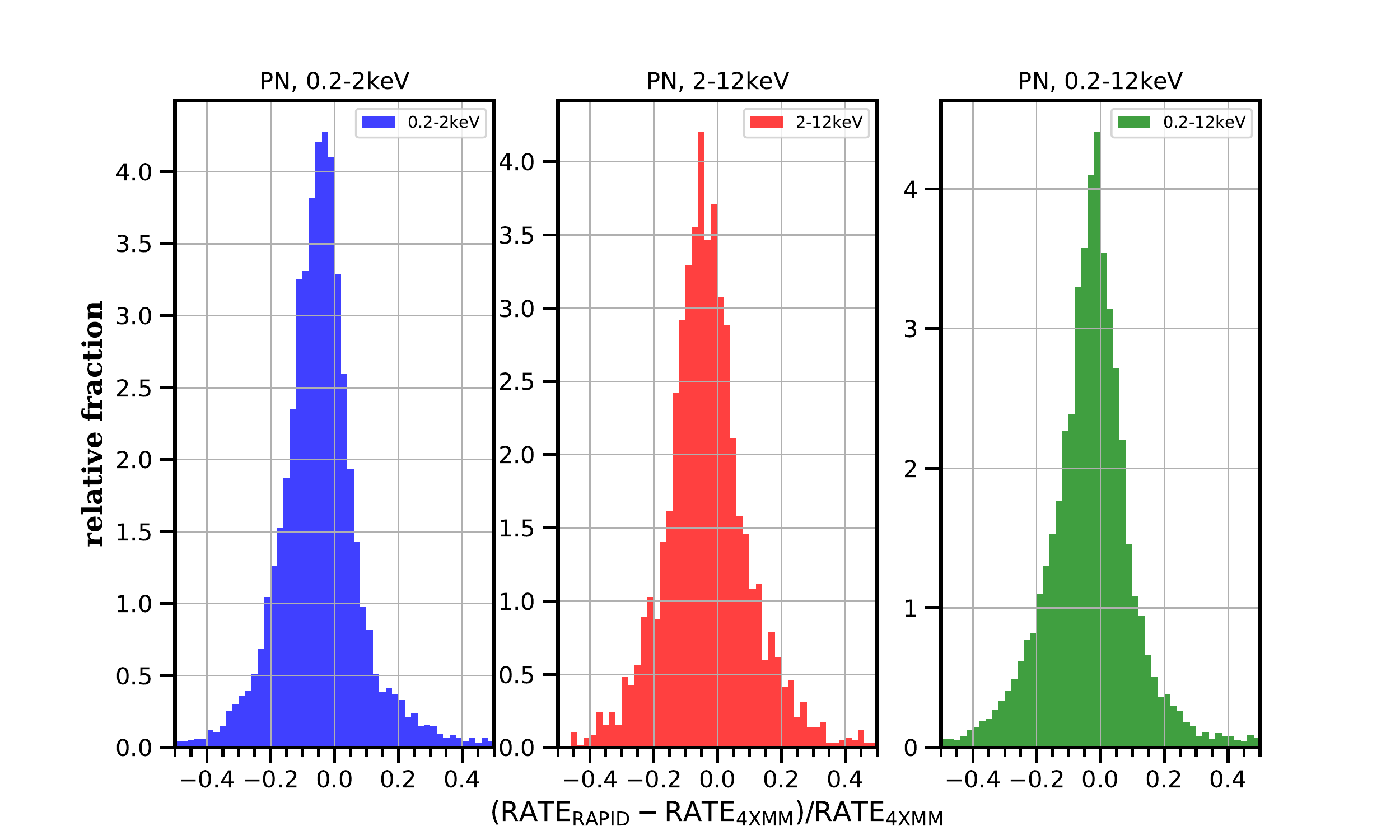}
\end{center}
\caption{Comparison between the 4XMM-DR9 and RapidXMM count rates for the PN detector. Top set of panels: the large window plots the 4XMM-DR9 count-rate vs the RapidXMM one. The smaller window underneath the main one shows the count-rate fractional difference between the 4XMM-DR9 and RapidXMM as a function of the 4XMM-DR9 count-rate. Each datapoint corresponds to a unique X-ray detection selected from the 4XMM-DR9 catalogue. The different colours are for each of the three RapidXMM energy bands, 0.2--2\,keV (blue), 2--12\,keV (red) and 0.2--12\,keV (green). The red and green set of points have been offset by 1\,dex and --1\,dex in the vertical axis for clarity. The dotted lines show the one-to-one relation. Bottom set of panels: distribution of the fractional difference between the 4XMM-DR9 and RapidXMM count rates. Different energy bands are shown in separate panels.}\label{fig:4xmm_ff_pn}
\end{figure}

\begin{figure}
\begin{center}
\includegraphics[angle=0,height=.6\columnwidth]{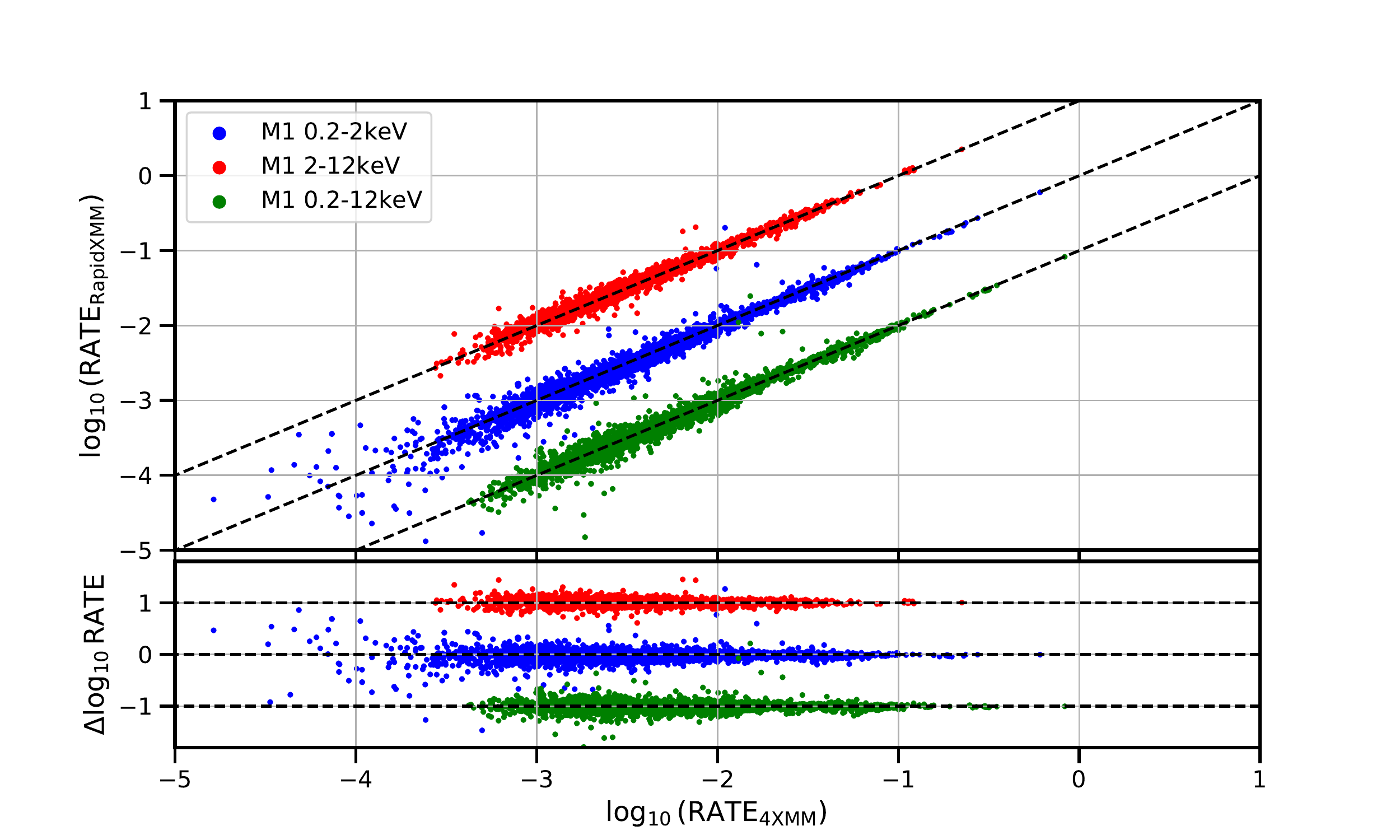}
\includegraphics[angle=0,height=.6\columnwidth]{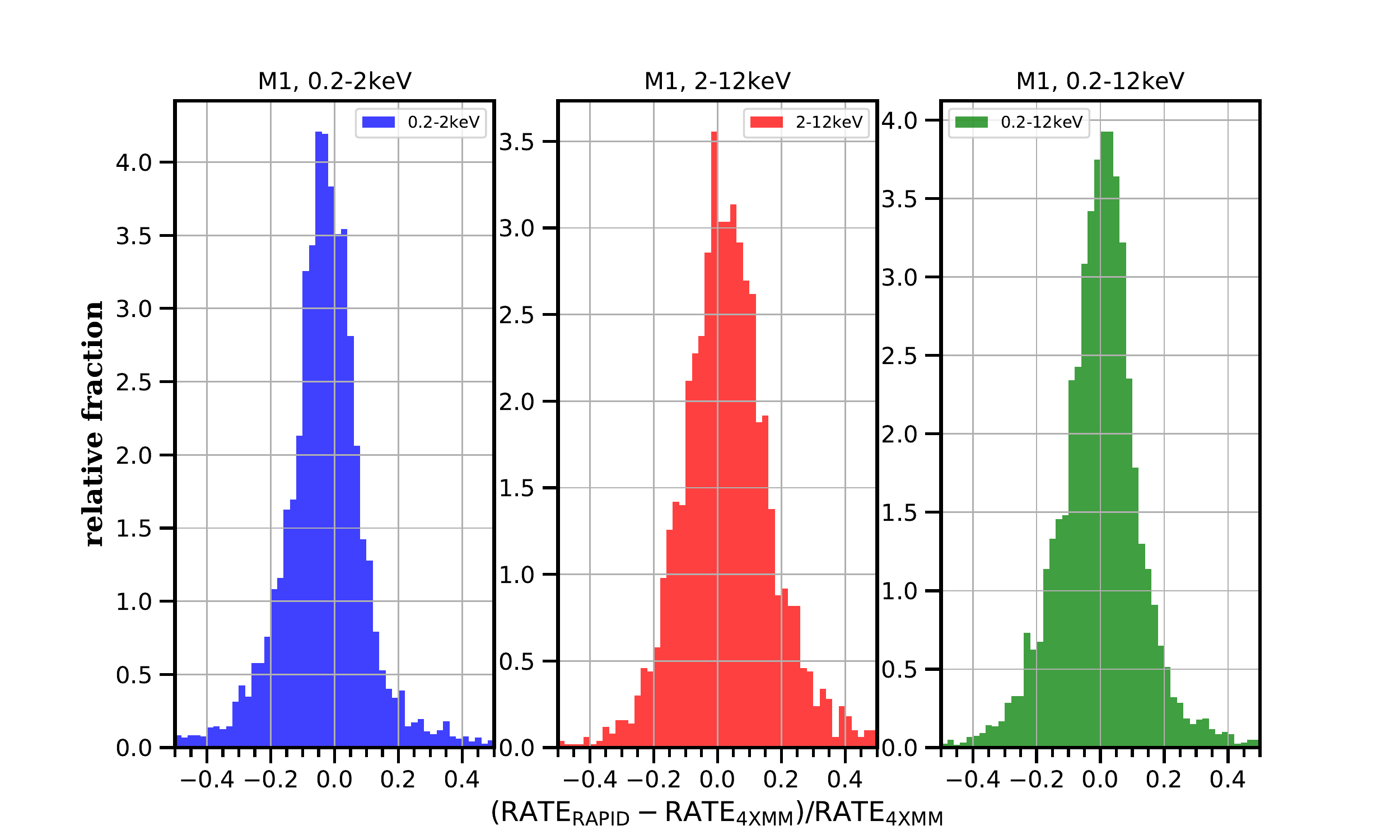}
\end{center}
\caption{Same as Fig.~\ref{fig:4xmm_ff_pn} for the EPIC-MOS1 camera.}\label{fig:4xmm_ff_m1}
\end{figure}

\subsection{HEALPix projection}
\label{sec:healpix-repro}
A key motivation of the RapidXMM project is to develop a methodology that enables the fast retrieval of upper limits at an arbitrary set of input sky coordinates. The first step to achieve this is to pre-calculate the upper limits within the \textit{XMM-Newton} footprint and store them in a database. This eliminates overheads associated with the real-time calculation of upper limits and enables nearest-neighbour queries to the database given a set of input positions. Such positional searches however, are cumbersome and therefore further optimisations are needed to facilitate the fast retrieval of the stored upper limits. The Hierarchical Equal Area isoLatitude Pixelisation \citep[HEALPix,][]{Gorski2005} of the sky is used to address this issue. The HEALPix scheme splits the 2-dimensional surface of the sphere into equal area cells that are uniquely identifiable via an integer number.\footnote{RapidXMM uses the HEALPix nested numbering scheme.} There is a one-to-one correspondence between the (sky) position of the cell on the sphere and its integer number and therefore, nearest-neighbour searches using sky coordinates are reduced to an integer matching exercise. This approach requires that the estimated upper limits are resampled to the HEALPix grid of cells. An important parameter in this exercise is the spatial resolution of the HEALPix tessellation that defines the size of cells. The decision on the HEALPix resolution is a trade-off between accurately describing spatial variations across the sky and size of the final database. The HEALPix {\sc order} parameter controls the resolution of the grid. A value of {\sc order=16} is adopted for Pointed observations and 15 for the Slew survey. These correspond to cell sizes of about 3.2 and 6.4\,arcsec respectively and are comparable to the pixel size of the \textit{XMM-Newton} PPS images ($\sim4$\,arcsec). We adopt a lower resolution for the Slew survey because of the low background of this dataset, which results into a more homogeneous spatial distribution of upper limits compared to Pointed observations. 

\begin{figure}
\begin{center}
\includegraphics[angle=0,height=.6\columnwidth]{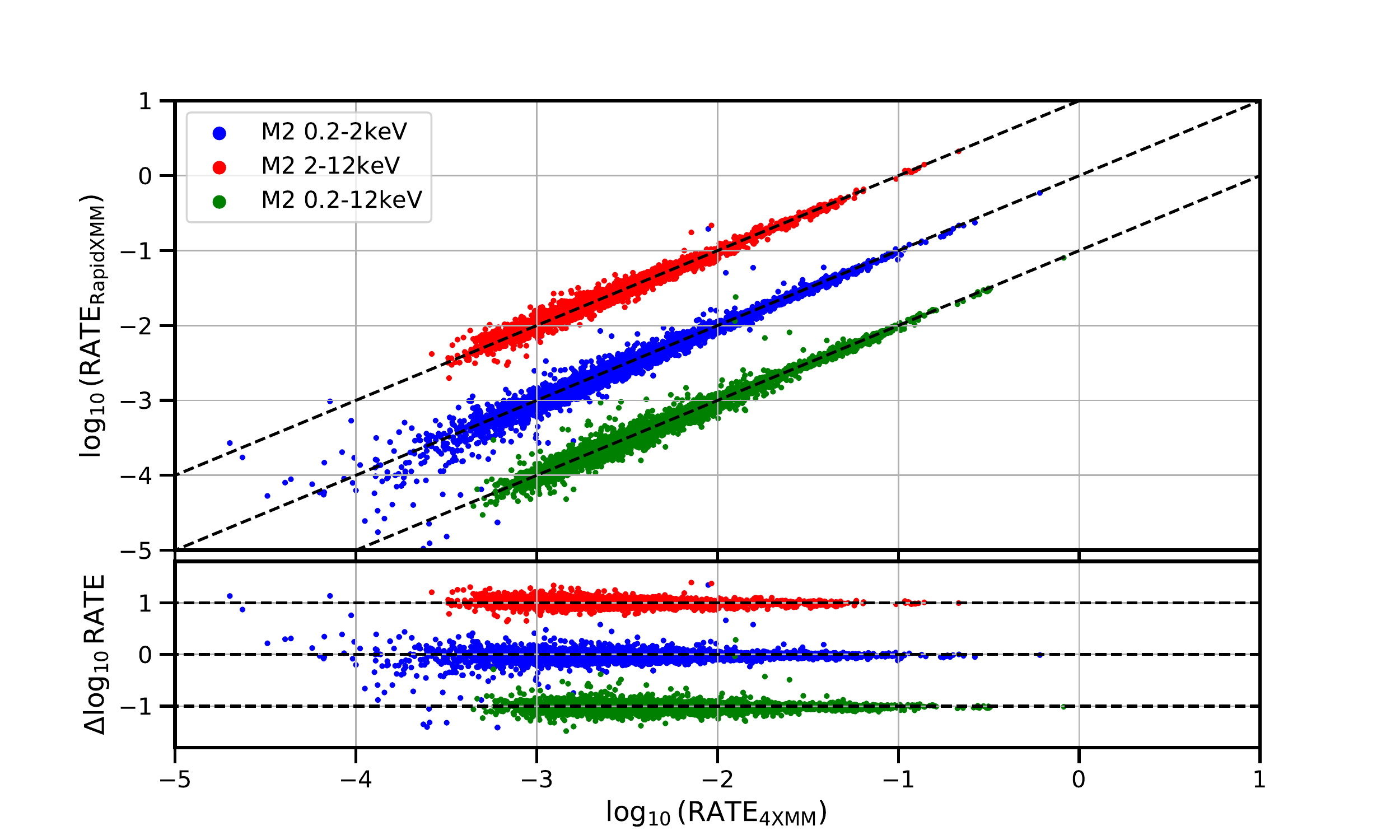}
\includegraphics[angle=0,height=.6\columnwidth]{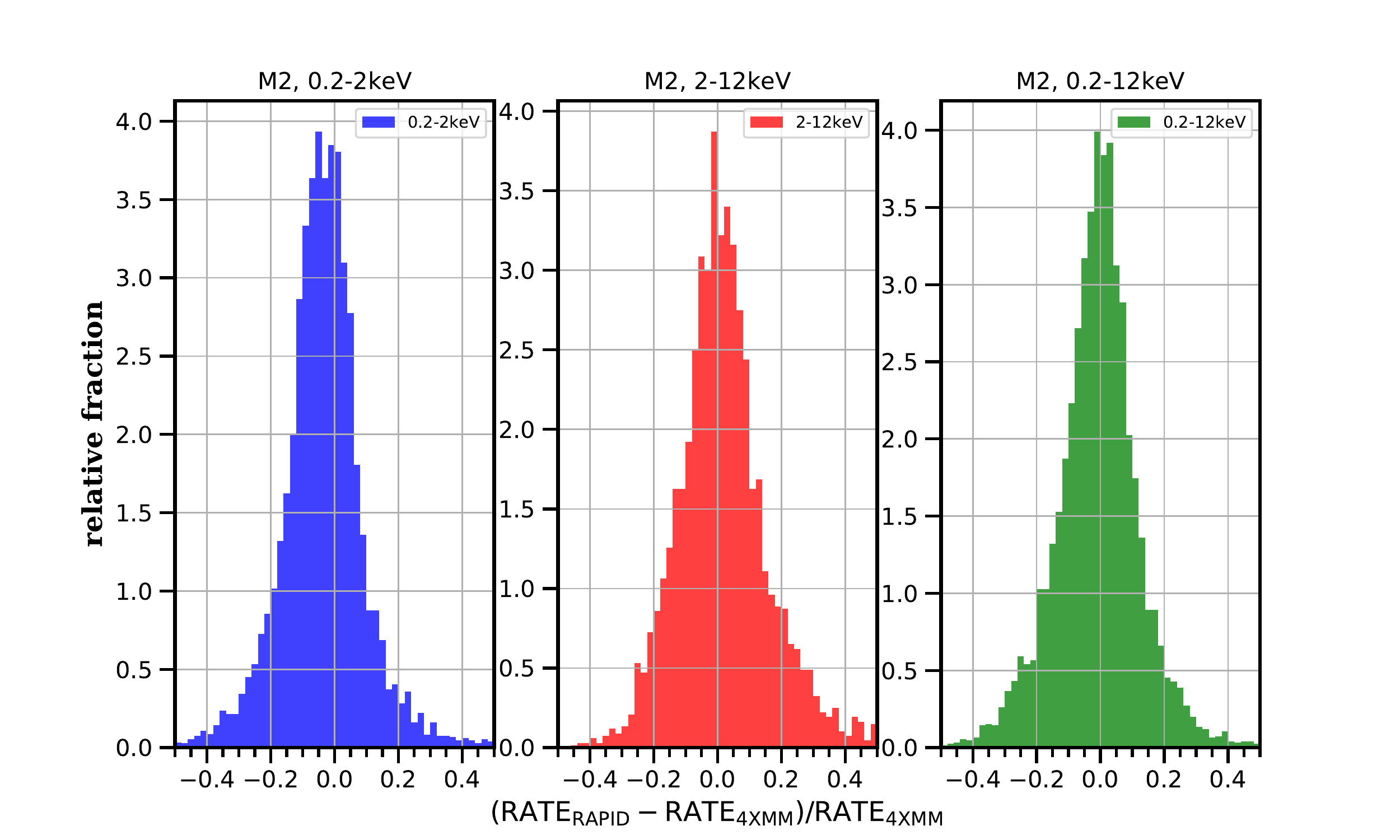}
\end{center}
\caption{Same as Fig.~\ref{fig:4xmm_ff_pn} for the EPIC-MOS2 camera.}\label{fig:4xmm_ff_m2}
\end{figure}

Next we describe how the data products generated in Sect.~\ref{sec:calc} are projected onto the HEALPix grid. The output of the upper-limit calculation stage are 2-dimensional images of upper limits and associated products (integrated counts, background expectations levels, mean exposure times, quality masks) in different energy bands and for each exposure or patch within an Obs.ID. Each of these images has the same pixel scale as the input PPS products (typically $\sim4$\,arcsec). Each HEALPix cell is assigned the upper limits (1, 2, 3$\sigma$) and associated products (source counts, background counts, average exposure time, area ratio and flags) of the image-pixel that lies nearest to each centre. Cells assigned to non-exposed pixels are discarded. In the case of Slew survey, there is significant overlap between patches within the same slew and therefore duplicates have to be removed to produce a unique list of HEALPix cells. Among each group of repeating cells in overlap regions the one with the best quality flag or those with the highest background counts are kept. Differences in the RapidXMM photometric products of a given HEALPix cell among  different patches of the same slew are to be expected. These are because of the different position of the HEALPix cell relative to the borders of the patches it lies on. This translates to variations in the effective extraction regions, and hence the estimated photometric quantities.

The final product of the HEALPix projection stage is a list of unique HEALPix cells for each exposure in an Obs.ID. Each of these cells is associated with upper limits and the products generated during the calculation stage (e.g. extracted counts, background expectation levels, mean exposure times, flags) at different energy bands. The list of quantities assigned to each cell are listed in Appendix~\ref{sec:db-fields}. Meta-data associated with the exposure or patch, such as Obs.ID number, \textit{XMM-Newton} observation mode, EPIC detector and filter, observation date, are also generated.

\subsection{The RapidXMM database}
\label{sec:rapidxmm-db}
The final products of Sect.~\ref{sec:healpix-repro}, i.e. list of HEALPix cells with their corresponding upper limits and ancillary data, are stored in the RapidXMM database implemented in a PostgreSQL 12.0 server. The Pointed and Slew survey data are stored in distinct tables with identical structure. The Schema of these tables is presented in detail in Appendix~\ref{sec:db-fields}. By using this simple database structure we minimise the use of joint queries and reduce the search response time. As explained in Sect.~\ref{sec:healpix-repro} above, the use of the HEALPix technology converts nearest-neighbour searches by sky-coordinate into an integer-matching exercise of the corresponding HEALPix cell numbers. The fields containing these integer cell numbers ({\sc npixel}) are indexed (i.e. an specific look-up table is generated for the field). This is a common technique in database design to speed-up queries using a selected field when the table has a very large number of entries. The RapidXMM database is accessible from the XMM-Newton Science Archive. More details can be found in Section \ref{sec:xsa}.

\begin{figure*}
\begin{center}
\includegraphics[angle=0,width=\textwidth]{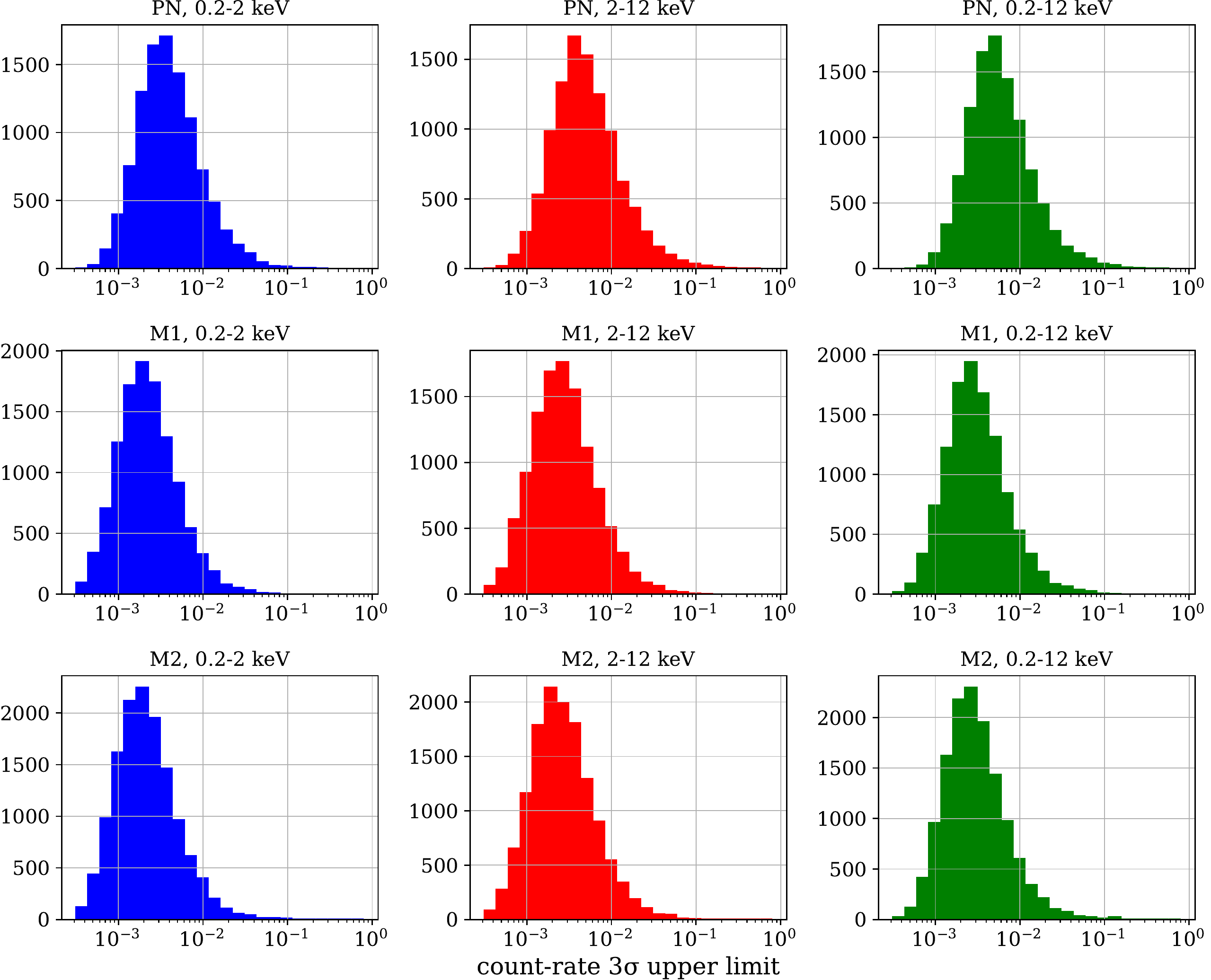}
\end{center}
\caption{Distribution of count-rate $3\sigma$ upper-limits for the EPIC-PN (top row), EPIC-MOS1 (middle row) and EPIC-MOS2 (bottom row) detectors. The left column shows the 0.2--2~keV energy band, the middle column corresponds to the 2--12~keV band and the right column is for the 0.2--12~keV band. These histograms are based on the upper limits returned by the RapidXMM database for a total of 10\,000 random sky positions within the 4XMM-DR9 footprint.}\label{fig:4xmm_crhist}
\end{figure*}

\subsection{Comparison with the {\sc eupper} task of SAS}
\label{sec:eupper}
The RapidXMM methodology for estimating upper limits based on aperture photometry is similar to the functionality offered by the {\sc eupper} task of SAS, which is currently used by HILIGT \citep{Saxton2022, Koenig2022}. Nevertheless, the RapidXMM analysis differs from that of  {\sc eupper} in a number of important ways. Non-exposed areas of the {\it XMM-Newton} field of view (e.g. CCD gaps, field edges) are accounted for in the RapidXMM analysis pipeline by measuring the fractional aperture area with positive exposure time. No such correction is applied by {\sc eupper}. RapidXMM assumes Poisson statistics for the estimation of upper limits (e.g. Equation \ref{eq:UL}) independent of the number of counts in the extraction aperture. The current implementation of {\sc eupper} (SAS version {\sc xmmsas\_20201028\_0905-19.0.0}) switches to a Gaussian approximation if the total counts in the extraction aperture exceed the value of 80. RapidXMM uses one-sided probabilities to define the $1\sigma$, $2\sigma$ and $3\sigma$ confidence intervals as opposed to two-sided for the {\sc eupper} task. For input positions in the vicinity of X-ray detections RapidXMM does {\it not} return upper limits, only photometry products, e.g. total counts, backgrounds, exposure time.

\section{Quality assessment and upper-limit statistical properties}
\label{sec:qa}

This section presents the quality assessment and the statistical properties of the products stored in the RapidXMM database, such as upper limits, extracted image counts, estimated background levels and exposure times. 

\begin{figure*}
\begin{center}
\includegraphics[angle=0,width=\textwidth]{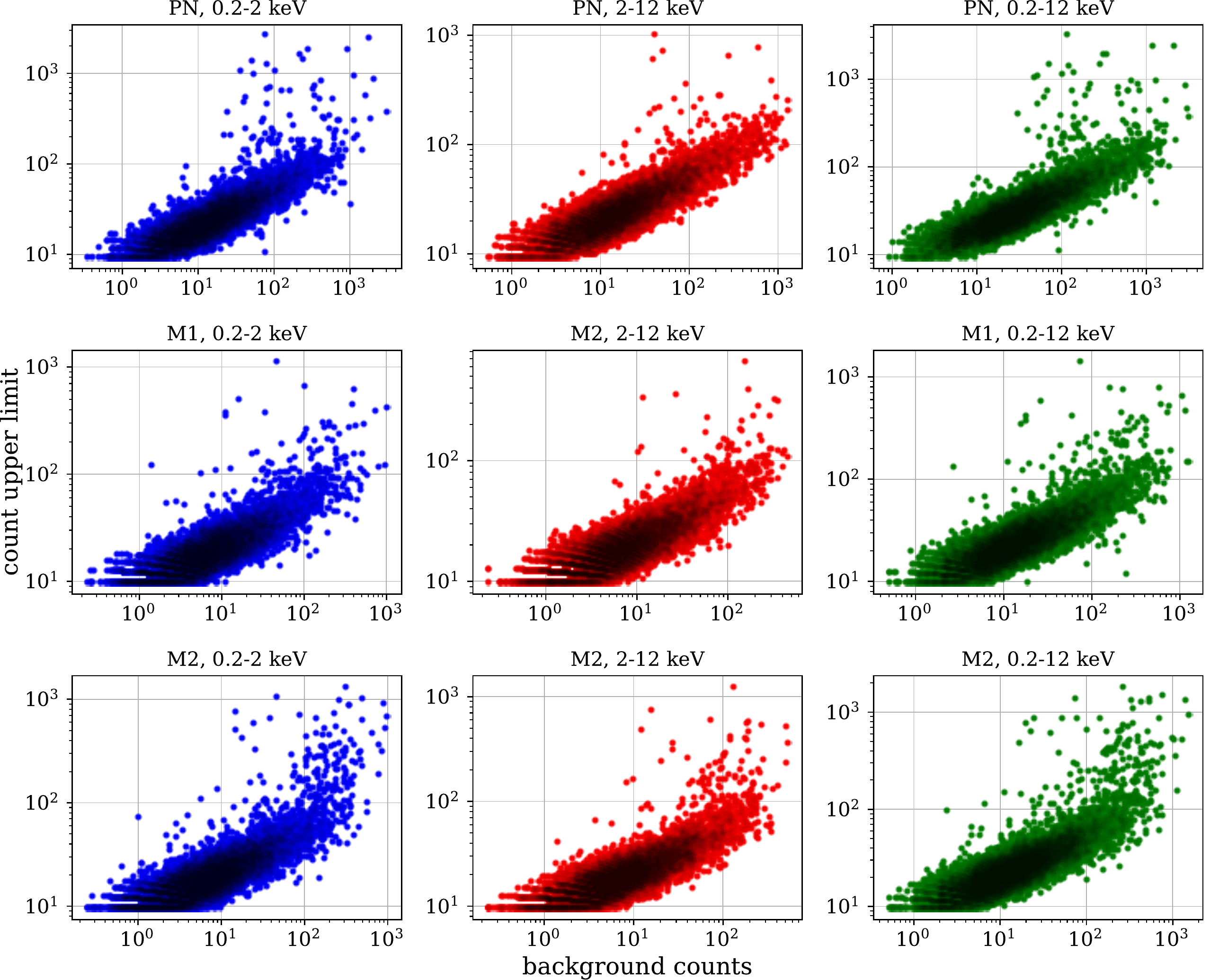}
\end{center}
\caption{$3\sigma$ count upper-limit as a function of the background level. The count upper-limits are estimated by multiplying the count-rate upper-limits returned by the RapidXMM database with the corresponding exposure time. The panels are as in Fig.~\ref{fig:4xmm_crhist}.}\label{fig:4xmm_culbkg}
\end{figure*}

First we test the X-ray photometry capabilities of  RapidXMM.  The count-rates of known X-ray sources are estimated using the RapidXMM products and then compared with those determined from external X-ray catalogues. This exercise does not directly assess the quality of the RapidXMM upper limits, but tests the consistency of the components that enter into the upper limit calculation (e.g. extracted counts, background levels, mean exposure times) with independent estimates. In the case of Pointed Observations the adopted external catalogue is the 4XMM-DR9 \citep{Webb2020}. For the Slew survey we use the 2nd XMM Slew Survey Catalogue (XMMSL2),\footnote{\url{www.cosmos.esa.int/web/XMM-Newton/xmmsl2-ug}} which is produced following the methods described in \cite{Saxton2008}. The second set of tests explores the overall statistical properties of the RapidXMM upper limits. This provides an overview of the dependence of the estimated upper limits on the background level, the exposure time and the {\it XMM-Newton} detector.

Because of differences in the analysis of Pointed and Slew Survey Observations, the quality assessment results for these two sets of data are presented separately.

\subsection{Pointed Observations}
In the case of Pointed Observations the photometry of the X-ray sources in the 4XMM-DR9 catalogue \citep{Webb2020} is compared to that determined independently by querying the RapidXMM database at the corresponding source positions. Each of the three EPIC cameras (PN, M1, M2) and each of the three RapidXMM energy bands (0.2--2, 2--12 and 0.2--12~keV) are independently analysed. The RapidXMM count rate at fixed energy band and for a given EPIC detector is
\begin{equation}\label{eq:CR_RAPID}
 {\rm RATE}_{\rm RapidXMM} = ( N - B ) / (t \cdot \mathrm{EEF}),
\end{equation}
where $N$ is the total number of counts within the RapidXMM extraction aperture (15\,arcsec for Pointed), $B$ is the expected background level, $t$ is the exposure time and EEF is the encircled energy fraction. Equation \ref{eq:CR_RAPID} provides an estimate of the sources' count rates in the energy intervals 0.2--2, 2--12 and 0.2--12\,keV. These should be compared to the 4XMM-DR9 photometry. The 0.2--12\,keV band is one of the standard photometry bands of the 4XMM-DR9 catalogue and therefore for this energy interval the 4XMM-DR9 and RapidXMM count-rates can be directly compared. This is not the case however, for the RapidXMM bands 0.2--2 and 2--12\,keV, for which no photometry is available in the 4XMM-DR9 catalogue. The 4XMM-DR9 count rates in these broad energy bands are instead synthesised by summing up the count rates in the 4XMM-DR9 narrow bands 1 (0.2--0.5\,keV), 2 (0.5--1.0\,keV), 3 (1.0--2.0\,keV), 4 (2.0--4.5\,keV) and 5 (4.5--12\,keV). For a given EPIC camera the rate in the 0.2--2\,keV energy interval is the sum of the 4XMM-DR9 rates in the narrow bands 1, 2 and 3. The 2--12\,keV rate is determined by summing the rates of the 4XMM-DR9 bands 4 and 5. Ideally the 4XMM-DR9 counts, background values and exposure times in each of the narrow bands above should be added/averaged to yield an estimate of the 4XMM-DR9 count rate of each source via Eq.~\ref{eq:CR_RAPID}. However, these quantities are not among those listed in the 4XMM-DR9 catalogue.  

\begin{figure*}
\begin{center}
\includegraphics[angle=0,width=\textwidth]{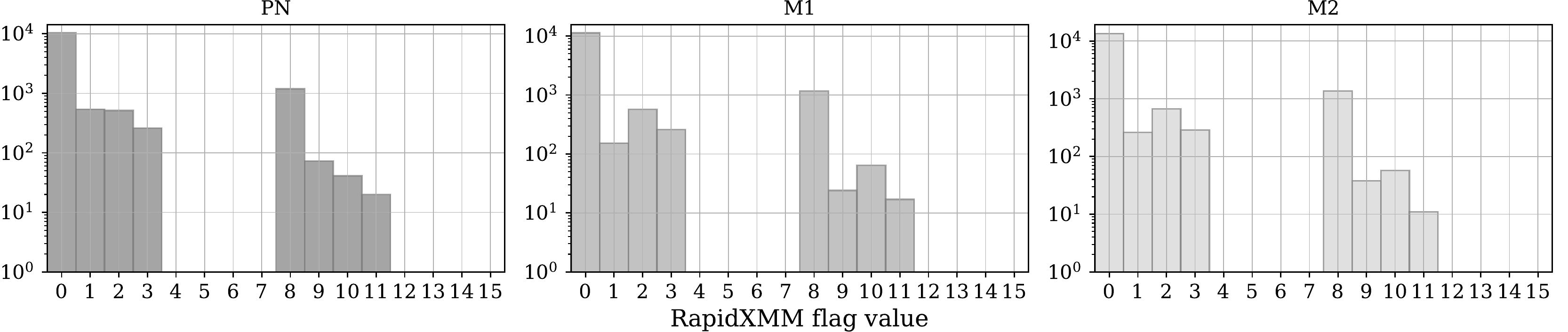}
\end{center}
\caption{Distribution of upper-limit quality flags for HEALPix cells on the EPIC-PN (left panel), EPIC-MOS1 (middle panel) and EPIC-MOS2 (right panel) detectors. These histograms have been constructed using the quality flags returned by the RapidXMM database for a total of 10\,000 random positions selected within the 4XMM-DR9 footprint.}\label{fig:4xmm_flaghist}
\end{figure*}

The photometry comparison between the RapidXMM and 4XMM-DR9 is limited to secure 4XMM-DR9 sources using the detection likelihood parameters listed in that catalogue. For the 2--12\,keV band, we apply a lower limit of 15 to the 4XMM-DR9 band 4 (2.0--4.5\,keV) detection likelihood parameter ($\rm \{\$ca\}\_4\_DET\_ML>15$, where $\rm \{\$ca\}$ takes values PN, M1, M2 for each of the EPIC cameras) to select secure hard-band sources. In the case of the 0.2--2 and 0.2--12\,keV bands the same threshold as above is applied to the band 8 (0.2--12\,keV) detection likelihood, i.e. $\rm \{\$ca\}\_8\_DET\_ML>15$. We further limit the count-rate comparison to 4XMM-DR9 sources that (i) are not X-ray extended by requiring that the 4XMM parameter $\rm EP\_EXTENT\_ML = 0$, (ii) lie at off-axis angles smaller than 13\,arcmin to avoid edge effects by thresholding the  4XMM-DR9 parameter \{\$ca\}\_OFFAXIS ($\rm \{\$ca\}$ as above), (iii) are not flagged by the PPS detection chain by requiring that the 4XMM-DR9 parameter SUM\_FLAG is zero, (iv) avoid sources with low detector coverage (i.e. with significant overlap with masked pixels) by requiring that the 4XMM-DR9 parameter is $\rm \{\$ca\}\_MASKFRAC>0.8$ (PSF weighted mean of the  coverage of a detection as derived from the detection mask).

In practice a total of 10\,000 4XMM-DR9 sources out of a total of 514\,870 that fulfil the above criteria are randomly selected from the catalogue. The RapidXMM database is queried at these 10,000 positions and the returned values are used to determine the count rates for each EPIC camera and each of the three RapidXMM energy bands using Eq.~\ref{eq:CR_RAPID}. A unique 4XMM-DR9 source may be detected in more than one \textit{XMM-Newton} observation, in which case the query to the RapidXMM database returns multiple products for each Obs.ID. The comparison with the 4XMM-DR9 count rates is limited to the same Obs.ID to avoid source variability issues. 

The results of the photometry comparison between RapidXMM and 4XMM-DR9 are presented in Figs.~\ref{fig:4xmm_ff_pn}, \ref{fig:4xmm_ff_m1}, \ref{fig:4xmm_ff_m2} for the detectors PN, M1 and M2 respectively. The top-panel of these figures plots the 4XMM-DR9 against the RapidXMM Count Rates in the 3 energy bands 0.2--2keV (blue), 2--12keV (red) and 0.2--12keV (green). The bottom panels show the distribution of fractional difference of the two count-rates for each energy band. Overall there is good agreement between the two independent estimates.  There are small systematic offsets of about 5\% or less. The effect is more pronounced in the 0.2--2 and 2--12keV bands. This is primarily related to the way these count-rates are estimated. The 4XMM fluxes are the sum of the count-rates in the narrow PPS energy bands, whereas the RapidXMM ones are the sum of the counts in these bands divided by the mean exposure time. Differences in the adopted EEFs and the overall flux estimation strategy (e.g. EMLDET vs aperture photometry) also contribute to this systematic offset.  The increased scatter in the CR vs CR plot for the 0.2--2keV band (blue) is because of the sample selection, which uses a detection likelihood threshold in the total 0.2--12\,keV band, $\rm \{\$ca\}\_8\_DET\_ML>15$, where $\rm \{\$ca\}$ is one of PN, M1, M2. This allows the inclusion in the sample of sources with lower significance in the 0.2--2\,keV band and hence higher Poisson uncertainties. Overall Figs.~\ref{fig:4xmm_ff_pn}, \ref{fig:4xmm_ff_m1}, \ref{fig:4xmm_ff_m2} show that the aperture-extracted counts, backgrounds and exposure times are consistent with the 4XMM-DR9 catalogue and provide a strong sanity check of the RapidXMM products. 

Next we explore the statistical properties of the upper limits estimated by the RapidXMM machinery. The 4XMM-DR9 Multi-Order-Coverage (MOC\footnote{\url{http://xmmssc.irap.omp.eu/Catalogue/4XMM-DR9/dr4moc.fits}}) map is used to randomly select a total of 10,000 positions within the sky footprint of the \textit{XMM-Newton} pointed observations used in the 4XMM-DR9 catalogue. The RapidXMM database is then queried at these positions and the returned products are analysed. The plots that follow are based on a total of 42,647 query results. There are multiple products returned at each position because of multiple overlapping exposures (typically PN, M1, M2) and Obs.IDs for a given input position. 
Figure \ref{fig:4xmm_crhist} shows the distribution of the $3\sigma$ upper-limits for each energy band and EPIC detector. The width of these distributions is related to the range of exposure times of individual \textit{XMM-Newton} observations in the archive and the spread in the background level among them. We further explore the dependence of the upper limits on the background level in Figure~\ref{fig:4xmm_culbkg}. This is constructed by first factoring out the impact of the exposure time on the inferred upper limits. This is achieved by multiplying the count-rate upper limits of Fig.~\ref{fig:4xmm_crhist} with the corresponding exposure time to convert them to upper limits in units of counts. These are then plotted as a function of background  level in Fig.~\ref{fig:4xmm_culbkg}. As expected the higher the background the less sensitive the (count) upper limits. This figure also shows the diversity of background levels in the \textit{XMM-Newton} archive. Finally, Fig.~\ref{fig:4xmm_flaghist} shows the distribution of RapidXMM quality flags at the 10,000 positions for each of the EPIC detectors.  As a reminder  flag values in the range 1--3 identify pixels close to CCD gaps or the edge of the field-of-view. Flags $\ge 8$ identify overlap with detected sources.

\begin{figure}
\begin{center}
\includegraphics[angle=0,height=.6\columnwidth]{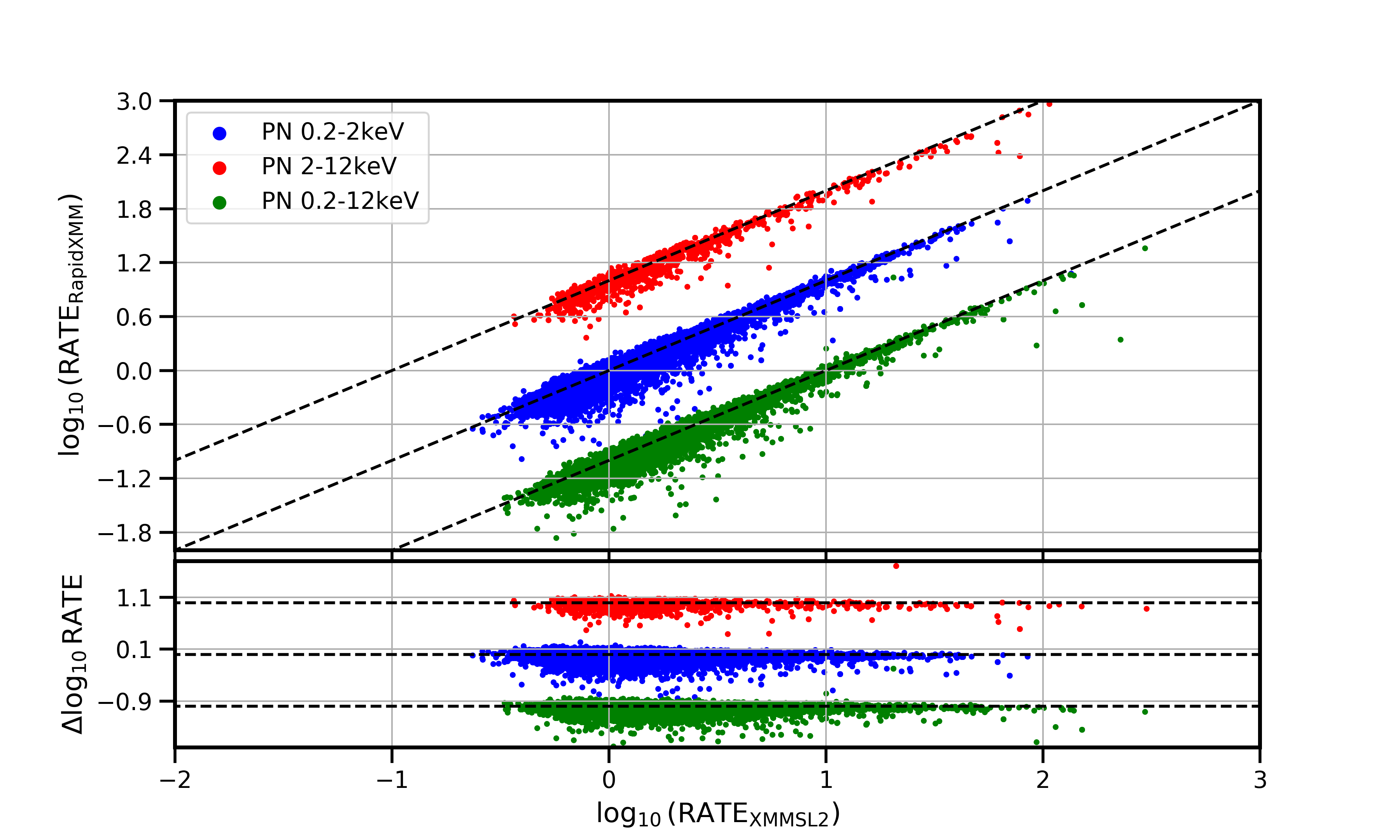}
\includegraphics[angle=0,height=.6\columnwidth]{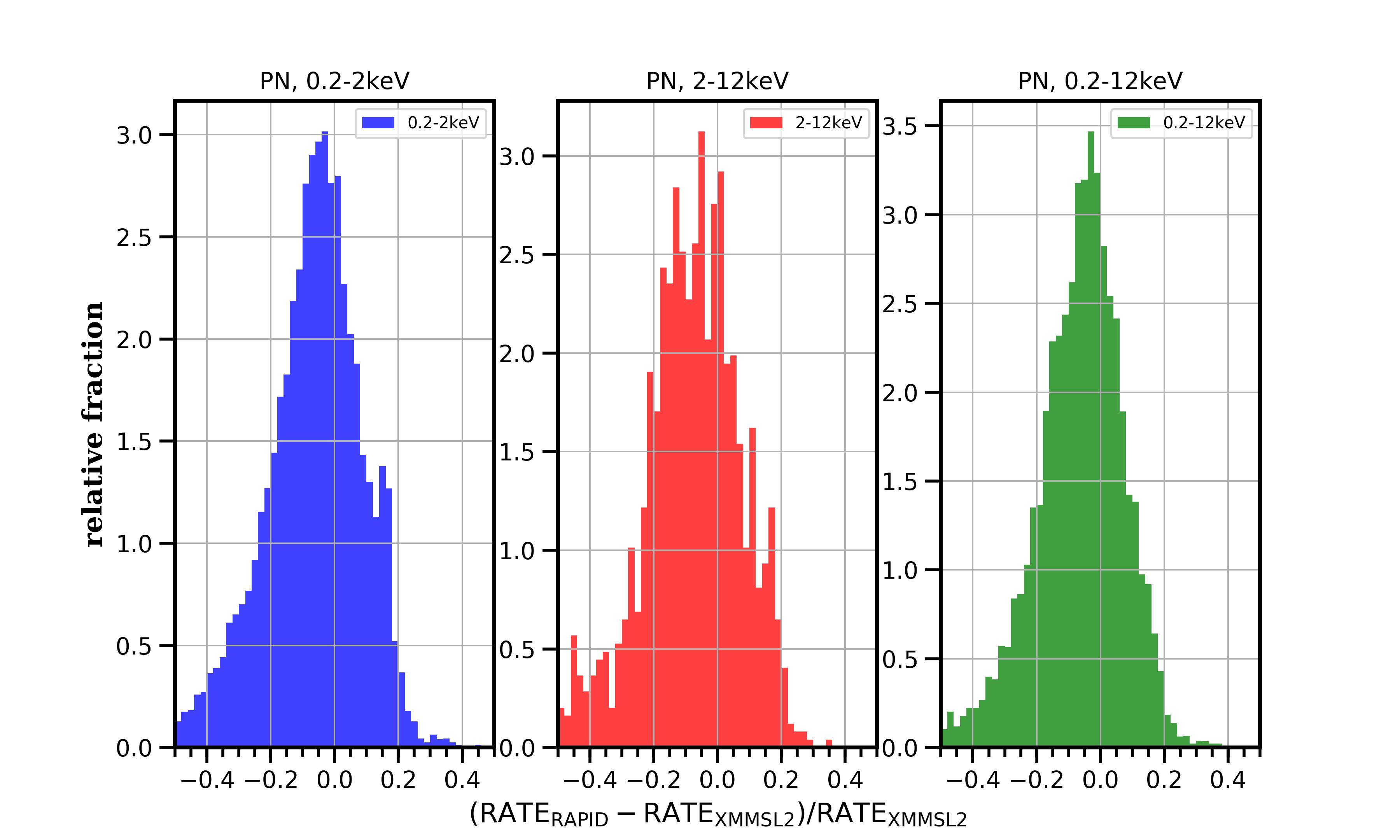}
\end{center}
\caption{Comparison between the XMMSL2 and RapidXMM count rates. Top set of panels: XMMSL2 source count-rate vs RapidXMM count-rate at the sources’ positions. Each datapoint corresponds to a unique X-ray detection in the XMMSL2 catalogue. The different colours correspond to the energy bands, 0.2--2\,keV (blue), 2--12\,keV (red) and 0.2--12\,keV (green). The red and green set of points have been offset by 1\,dex and --1\,dex in the vertical axis for clarity. The dotted lines show the 1--1 relation. Bottom set of panels: distribution of the fractional difference between the XMMSL2 and RapidXMM count rates. Different energy bands are shown in separate panels.}\label{fig:xmmsl_ff_pn}
\end{figure}

\begin{figure*}
\begin{center}
\includegraphics[angle=0,width=\textwidth]{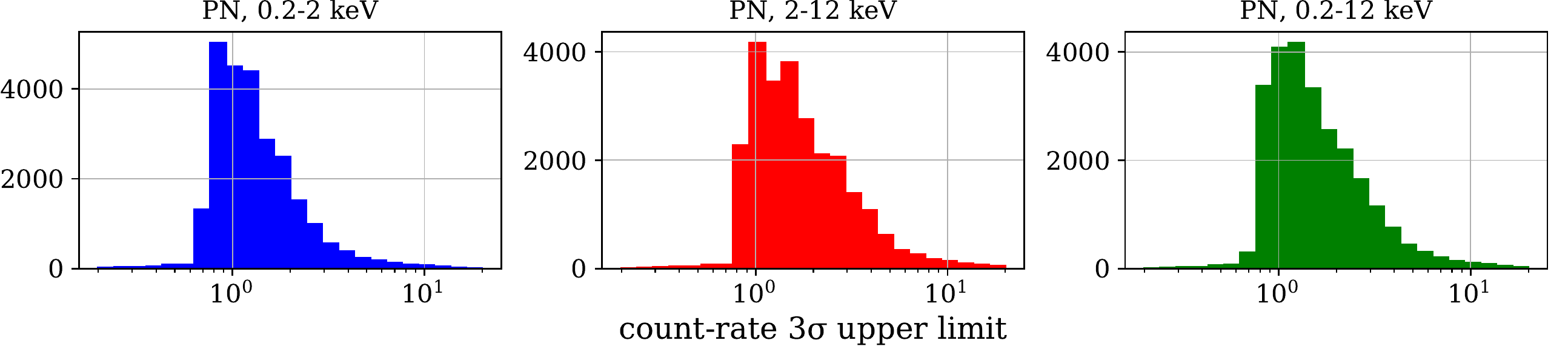}
\end{center}
\caption{Distribution of count-rate $3\sigma$ upper limits for the Slew Survey. The left column shows the 0.2--2~keV energy band, the middle column shows the 2--12~keV band and the right column show the 0.2--12~keV band. The histograms are constructed using the upper limits returned by the RapidXMM database for a total of 10\,000 random positions distributed across the sky.}\label{fig:xmmsl_crhist}
\end{figure*}

\begin{figure*}
\begin{center}
\includegraphics[angle=0,width=\textwidth]{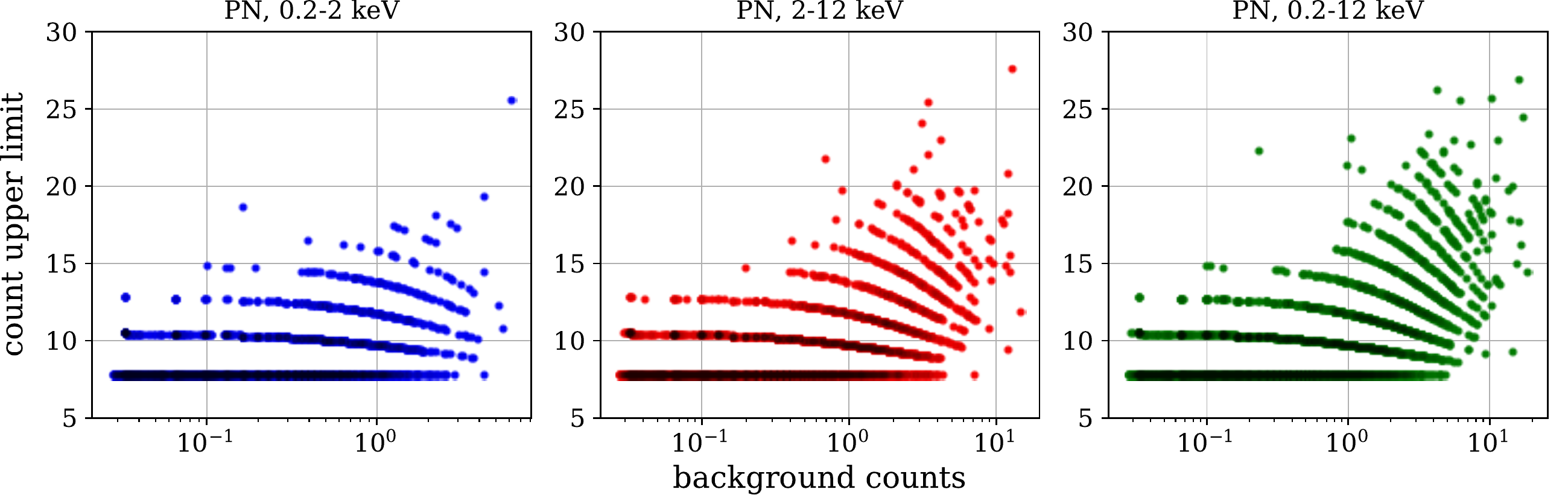}
\end{center}
\caption{$3\sigma$ count upper limit for the Slew Survey as a function of the background level. The count upper limit is estimated by multiplying the count-rate upper limit returned by the RapidXMM database with the corresponding exposure time. The data points correspond to the upper limits returned by the RapidXMM database for a total of 10\,000 random positions distributed across the sky. The panels are as in Fig.~\ref{fig:xmmsl_crhist}.}\label{fig::xmmsl_culbkg}
\end{figure*}

\subsection{Slew Survey Observations}
In the case of Slew Survey observations the count rates of X-ray sources in the XMMSL2 catalogue are compared against those determined independently by querying the RapidXMM database at the corresponding source positions. The RapidXMM photometry in the energy intervals 0.2--2, 2--12 and 0.2--12\,keV is estimated via Eq.~\ref{eq:CR_RAPID}. The XMMSL2 catalogue provides count rates in the same spectral bands. The corresponding catalogue parameters are RATE\_B6 (0.2--2\,keV), RATE\_B7 (2--12\,keV) and RATE\_B8 (0.2--12\,keV). 

We limit the comparison to XMMSL2 sources (i) with detection likelihood (catalogue columns DET\_ML\_BAND\{\$EN\}, \{\$EN\} = 6, 7, 8) $>20$ in the energy band of interest, (ii) are not extended (catalogue columns EXT\_B\{\$EN\}==0) and (iii) are not flagged by the detection chain by requiring that the logical columns VER\_HALO, VER\_HIBGND, VER\_PSUSP and VER\_FALSE are all set to false. A total of 14,588 XMMSL2 sources fulfil the above criteria. The RapidXMM database is queried at the sources' positions and the returned values (total of 72,332) are used to determine the count rates for each energy band using Eq.~\ref{eq:CR_RAPID}. A unique source may overlap with more than one Obs.IDs in which case multiple count-rates are estimated. The comparison with the XMMSL2 is limited to the same Obs.ID  to avoid source variability. 

The results of this comparison are presented in Figure \ref{fig:xmmsl_ff_pn}. The top-panel plots the XMMSL2 against the RapidXMM count rates in the three energy bands 0.2--2\,keV (blue), 2--12\,keV (red) and 0.2--12\,keV (green). The bottom panel shows the distribution of the fractional difference of the two count-rates for each energy band. Overall there is acceptable agreement between the two independent estimates although there is evidence in Fig.~\ref{fig:xmmsl_ff_pn} that the RapidXMM count rates lie systematically below the XMMSL2 ones. This is manifested by a systematic offset of about 5--7\% of the peaks of the distributions in that figure toward negative values and the stronger tails at negative count-rate fractional difference.  We checked that this discrepancy is not related to differences in the background or exposure time estimation between the XMMSL2 and RapidXMM. Instead sources in the tails of the distributions of Figure \ref{fig:xmmsl_ff_pn} show a systematic offset in the total net counts determined by RapidXMM and the XMMSL2 analysis pipeline. Visual inspection of a sample of sources with discrepant photometry indicates that the differences are because of (i) residual artifacts on the Slew images, such as cosmic-ray hits, (ii) proximity to X-ray extended sources, (iii) the different methods of measuring count rates adopted by RapidXMM and the XMMSL2, i.e. aperture photometry versus PSF fitting.

\begin{figure}
\begin{center}
\includegraphics[angle=0,width=\columnwidth]{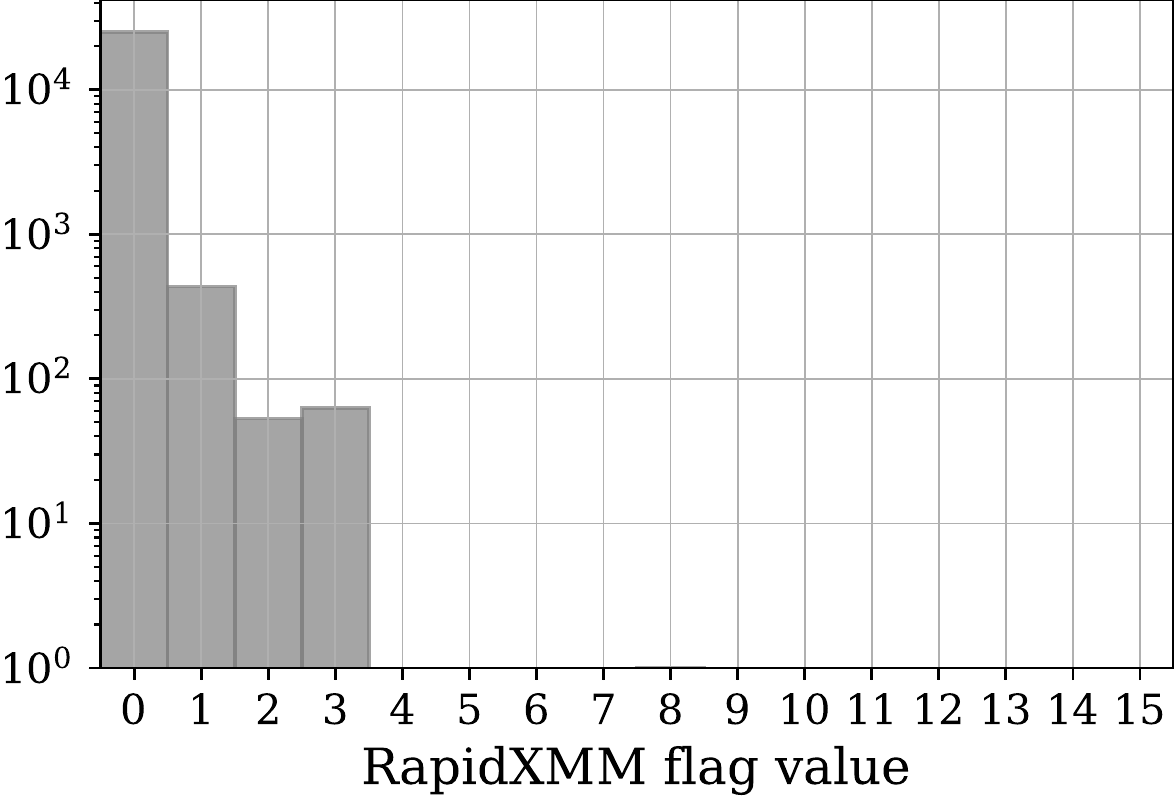}
\end{center}
\caption{Distribution of upper-limit quality flags for HEALPix cells in the Slew Survey.}\label{fig:xmmsl_flaghist}
\end{figure}

Next we explore the statistical properties of the Slew upper limits estimated by the RapidXMM machinery. A total of 10,000 random positions on the sky are generated. The RapidXMM database is then queried at these positions and the returned products are analysed. The plots that follow are based on a total of 25,693 query results. There are multiple products returned at each position because of overlapping Obs.ID footprints at a given input position. Figure~\ref{fig::xmmsl_culbkg} shows the count-rate $3\sigma$ upper limits at the queried positions at different bands. Because of the relatively homogeneous exposure time of the Slew observations and the very low background values the width of these distributions is primarily related to the number of extracted counts within the RapidXMM aperture. This is demonstrated in Fig.~\ref{fig::xmmsl_culbkg} which shows the $3\sigma$  count upper-limit (i.e. count-rate upper-limit multiplied by the corresponding exposure time) as a function of the background level. The stripes in this plot correspond to different integer number of extracted counts within the aperture. The lower stripe corresponds to zero counts. Figure \ref{fig:xmmsl_flaghist} shows the distribution of the RapidXMM quality flags in the case of the Slew survey observations. The density of sources in the Slew survey is low and that reflects on the small number of positions with flag values of 8 or greater. The vast majority of positive flags is related to the edge of field of view of the slew observations.  

\section{Science Applications}
\label{sec:science-applications}

This section presents two simple show-case examples of scientific applications using the RapidXMM database. In Sect.~\ref{sec:variability} we show how RapidXMM can be employed for the identification of extreme X-ray variability. In Sect.~\ref{sec:stacking} we use RapidXMM to perform stacking analysis of astrophysical populations that are not individually detected at X-rays.

\begin{figure}
\begin{center}
\includegraphics[angle=0,width=1\columnwidth]{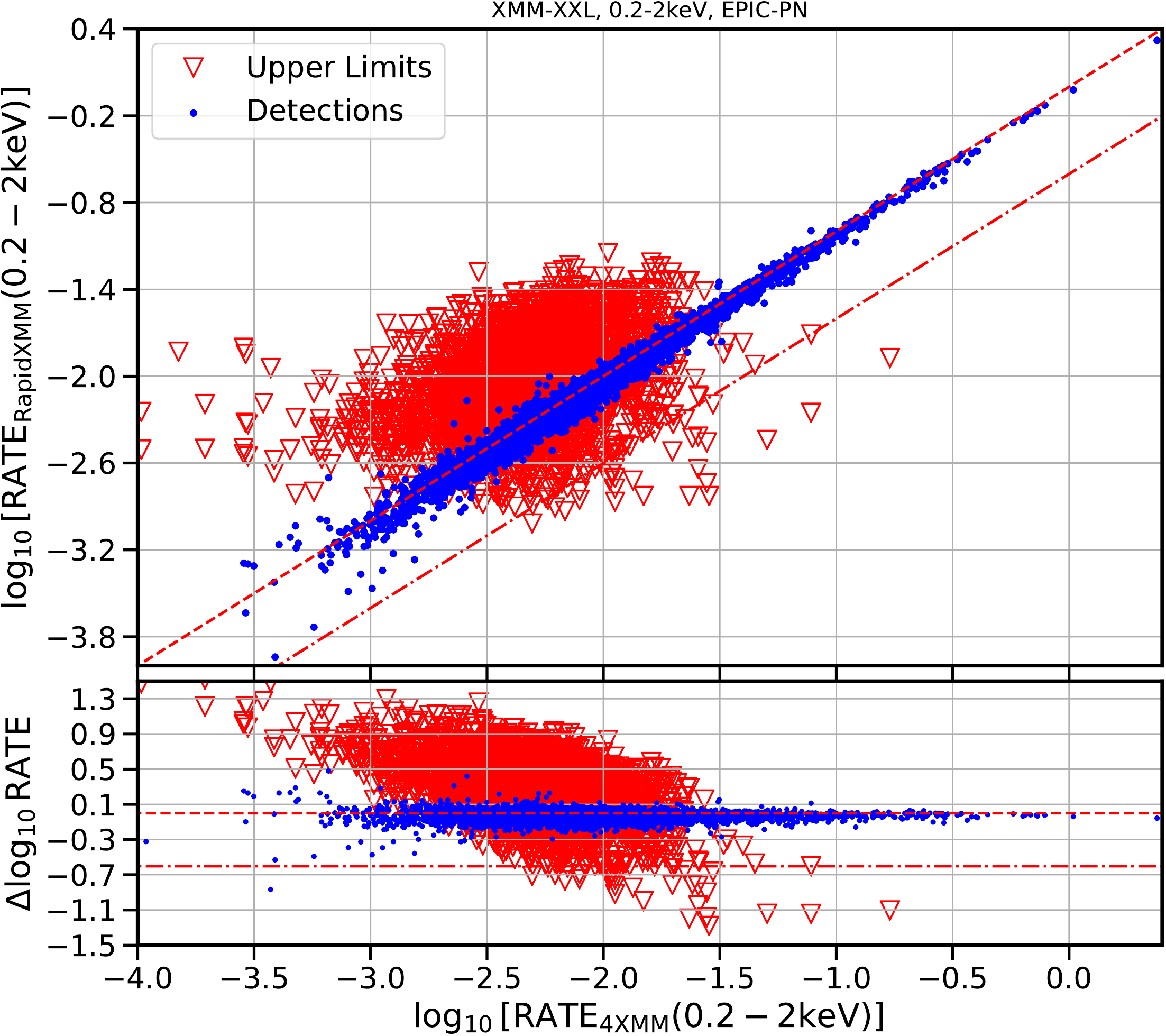}
\end{center}
\caption{Comparison between the 4XMM-DR9 and multi-epoch RapidXMM photometry for sources in the XMM-XXL (see Table~\ref{tab:xxl-sel}). The top panel plots the 4XMM-DR9 count-rate against the RapidXMM one. The blue dots correspond to the same Obs.ID as the one of the 4XMM-DR9 source. The red triangles are $3\sigma$ upper limits estimated for Obs.IDs on which the corresponding 4XMM-DR9 source is not formally detected. The short-dotted line shows the one-to-one count-rate relation. The dot-dashed line lies a factor of 4 below the one-to-one relation. The bottom panel plots the fractional difference between the 4XMM-DR9 and RapidXMM photometry as a function of 4XMM-DR9 count rate. The data point colours/shape and the lines have the same meaning as in the top panel.}\label{fig:xxl_ff}
\end{figure}

\begin{figure*}
\begin{minipage}[b]{0.49\linewidth}
\centering
\includegraphics[width=\textwidth]{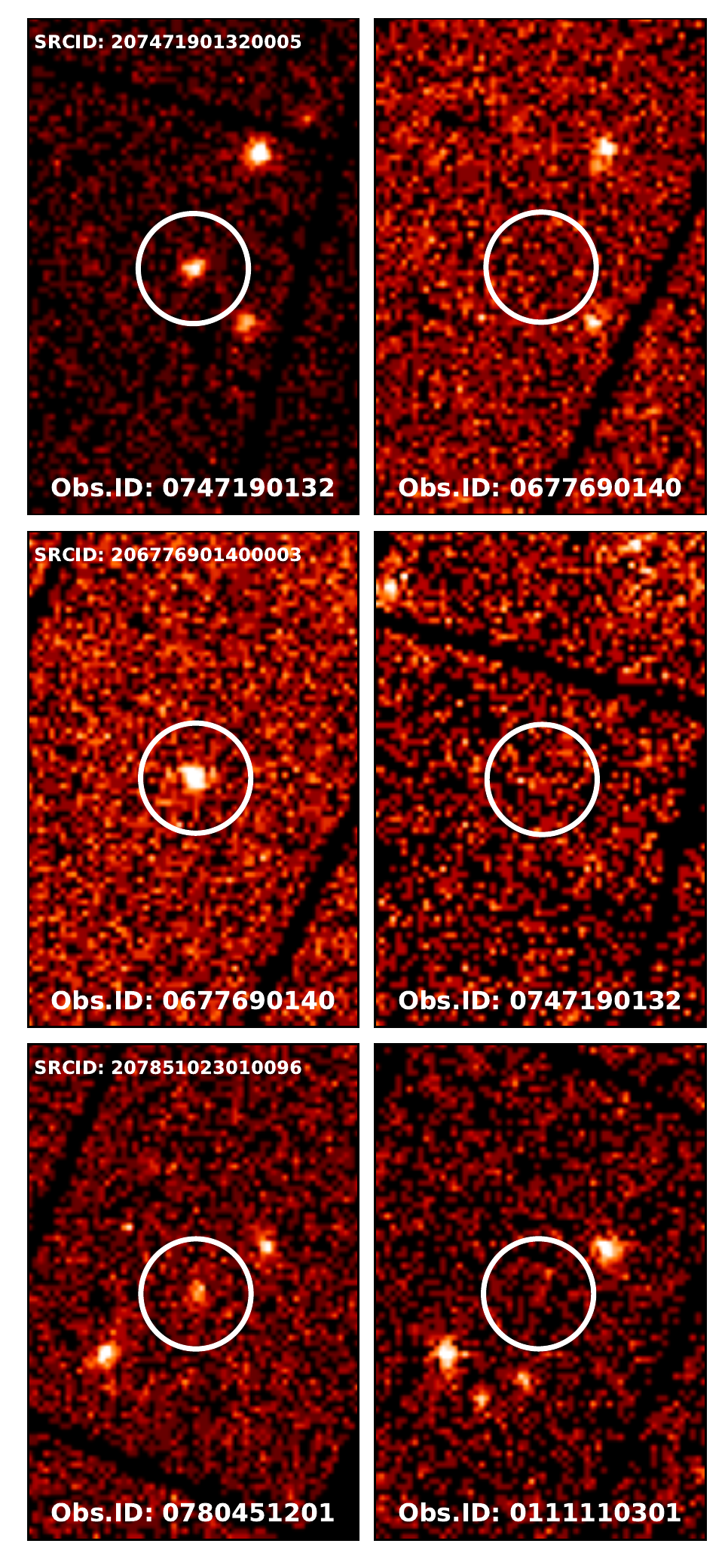}
\end{minipage}
\hspace{0.1cm}
\begin{minipage}[b]{0.49\linewidth}
\centering
\includegraphics[width=\textwidth]{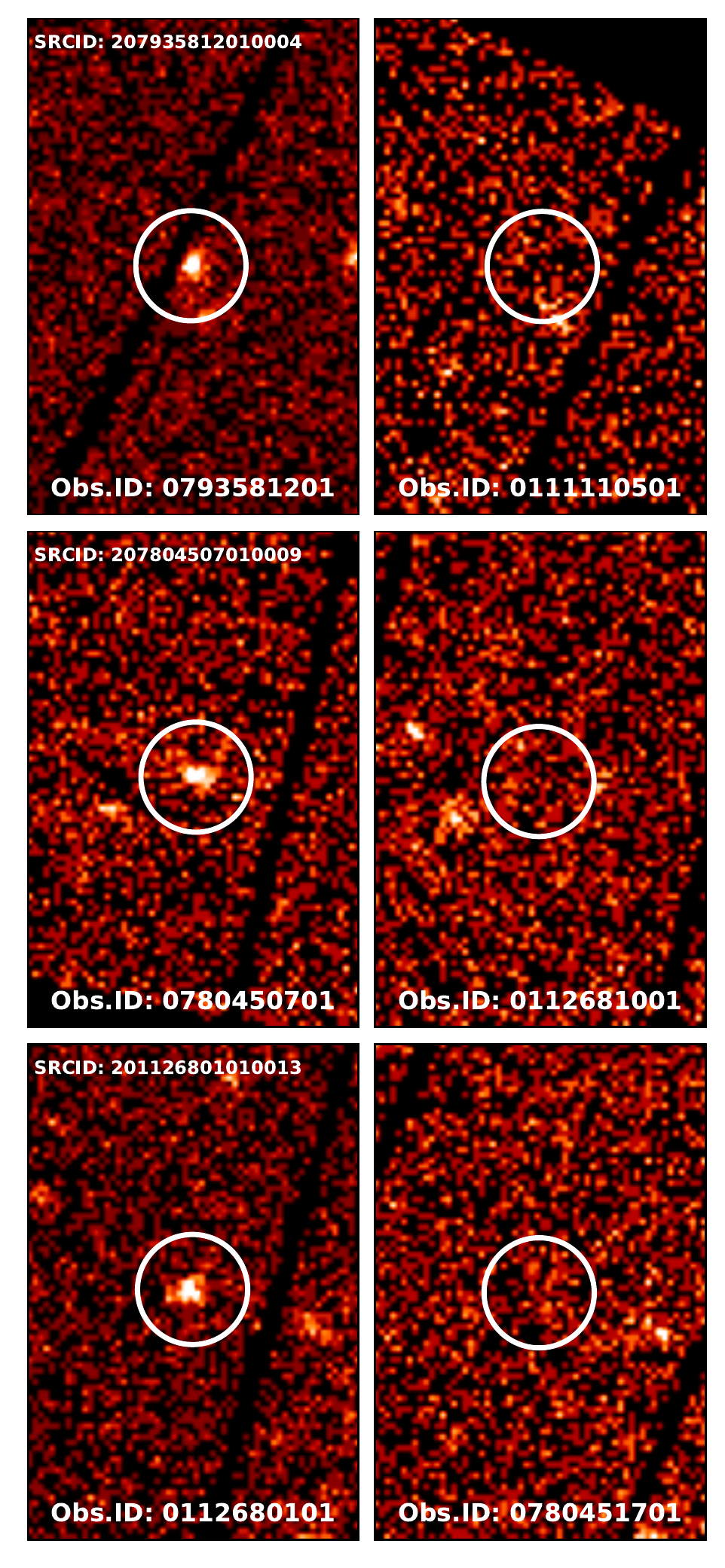}
\end{minipage}
\caption{X-ray thumbnails ($6\times4$ arcmin, north at the top, east to the left) of the vanishing X-ray sources listed in Table \ref{tab:xxl-sources}. Each pair of images corresponds to a single 4XMM-DR9 source. The images on the left correspond to the Obs.IDs on which the sources are detected. The images on the right correspond to the Obs.IDs on which the sources disappear. The Obs.ID is shown at the bottom on each thumbnail. The unique identification number (SRCID) of each source in the 4XMM-DR9 catalogue is shown on the images on the left. In all panels the circles are 40\,arcsec in radius. All images correspond to the 0.2--12~keV band.}
\label{fig:xxl-thumb}
\end{figure*}

\subsection{Vanishing X-ray sources}
\label{sec:variability}
The goal is to identify sources that undergo large-amplitude variations of their flux between different {\it XMM-Newton} observation epochs. We focus in particular on examples of flux suppression that make a given source disappear into the noise and below the formal detection limit of an observation. The choice of searches of such sources is that they provide an excellent demonstration of the transient X-ray sky that missions like eROSITA \citep{Merloni2012} will systematically explore. 

The RapidXMM database is queried at the positions of 4XMM-DR9 sources. Each input position may overlap with more than one \textit{XMM-Newton} pointings. In at least one of these pointings the sources are detected by the \textit{XMM-Newton} Processing Pipeline Subsystem (PPS) and are registered in the 4XMM-DR9 catalogue. A source however, may not be formally detected in all the observations it overlaps with. This may happen if the effective exposure time at the source position is low (e.g. shallow observations or large off-axis angles), if there are large variations in the background level of the overlapping \textit{XMM-Newton} pointings or if the source's flux varies substantially between observations. The latter is the focus of this section. It is possible to identify such cases by comparing the count-rate of unique sources in 4XMM-DR9  with the count-rate upper limits at the various epochs that the source positions under consideration have been observed by \textit{XMM-Newton}. If the upper limit of a given source lies significantly below the catalogued source count-rate then this is a candidate vanishing source.  

The RapidXMM infrastructure is well suited for this type of searches. The database contains integrated counts, mean exposure times and expected background levels within 15\,arcsec apertures in a grid of positions separated by about approximately 3\,arcsec. Quality flags are included to indicate if a position in the database overlaps with an X-ray source detected by the PPS detection chain, or if it lies close to the edge of the field of view of an \textit{XMM-Newton} pointing. In these cases no upper limits are determined. Overlapping observations are analysed independently thereby enabling access to multi-epoch information. It is therefore straightforward to carry out X-ray photometry at 4XMM-DR9 source positions and query upper limits, if available, in all the multi-epoch observations for a given sky position. 

\begin{table}
\caption{Selection criteria to identify 4XMM-DR9 sources that overlap with the XMM-XXL field and show large level of variability between observation epochs.}\label{tab:xxl-sel}
\begin{center}
\begin{tabular}{lll}
\hline
	4XMM-DR9 parameter & description & Number\\ 
	                   &           & of sources  \\
    \hline
 $30<$RA$<39\rm \,deg$ & Right Ascension  & \multirow{2}{*}{15101}\\
 $-7.5<$DEC$<-2.5\rm \,deg$ & Declination & \\
    PN\_8\_DET\_ML$>$20 & EPIC PN detection likelihood & 11440\\
    EP\_EXTENT\_ML$=$0 & EPIC extension likelihood & 10124\\
    PN\_OFFAX$<$13\,arcmin & EPIC-PN off-axis angle & 8759\\
\hline	 	 	 	 
\end{tabular}
\end{center}
\end{table}

\begin{table*}
\centering
\caption{List of XMM-XXL variable sources selected to have ratio between 4XMM-DR9 count rate and RapidXMM $3\sigma$ upper limit $>4$. The columns are (1) 4XMM-DR9 identification number, (2) Obs.ID on which the source is detected, (3) Right Ascension in J2000 of the detected X-ray source, (4) Declination in J2000 of the detected source, (5) Modified Julian Date of the \textit{XMM-Newton} observation, (6) logarithm base 10 of the 4XMM-DR9 count rate in the 0.2--2\,keV band of the source estimated by summing the count-rates in 4XMM-DR9 bands 1 (0.2--0.5\,keV), 2 (0.5--1\,keV) and 3 (1--2\,keV), (7) \textit{XMM-Newton} Obs.ID on which the source is not detected and an upper limit is estimated, (8) logarithm base 10 of the count rate $3\sigma$ upper limit in the 0.2--2\,keV band stored in the RapidXMM database, (9) Modified Julian Date of the \textit{XMM-Newton} observation to which the upper limit corresponds, (10) ratio between 4XMM-DR9 count rate and RapidXMM $3\sigma$ upper limit in the 0.2--2\,keV band.}
\label{tab:xxl-sources}
\begin{tabular}{ccc ccc ccc c}
\hline 
4XMM ID & Det. Obs.ID & RA & DEC & Det. MJD  & $\log_{10}\rm (Cnt\;Rate)$  &  UL Obs.ID & $\log_{10}(\rm Upper\; Limit)$ & UL MJD & ratio \\
        &             & (J2000) & (J2000)  & &  (cnt/s; 0.2--2\,keV)  &   & (cnt/s; 0.2--2\,keV) & &   \\
 (1)    &  (2)        & (3)     & (4)      & (5) & (6) & (7)  & (8) & (9) & (10) \\
\hline

207471901320005  & 0747190132 & 2:02:00.223 & $-$7:06:48.016 & 56687 & $-1.70$ & 0677690140 & $<-2.52$ & 55776 & $>6.53$ \\

206776901400003 & 0677690140 & 2:02:18.952 & $-$7:04:50.703 & 55775 & $-1.61$ & 0747190132 & $<-2.43$  & 56687 & $>6.44$\\

207851023010096 & 0780451201 & 2:25:41.988 & $-$5:11:17.468 & 57763 & $-2.15$ & 0111110301 & $<-2.77$ & 52093.3 & $>4.16$\\

207935812010004 & 0793581201 & 2:22:03.158 & $-$5:03:11.168 & 57754 & $-1.59$ &  0111110501 &  $<-2.64$ & 52094.3 & $>11.18$ \\

207804507010009$^1$ & 0780450701 & 2:25:41.672 & $-$4:34:16.748 & 57614 & $-1.63$ & 0112681001 & $<-2.83$ & 52304.7 & $>15.70$ \\

201126801010013$^2$ & 0112680101 & 2:27:07.275 & $-$4:04:36.667 & 52302 &  $-2.04$ & 0780451701 & $<-2.64$ & 57763.8 & $>4.00$ \\

\hline
\multicolumn{10}{l}{$^1$ Source is associated with a $z=0.601$ AGN} \\ 
\multicolumn{10}{l}{$^2$ Source is associated with a $z=0.492$ AGN}
\end{tabular}
\label{tab:variability}
\end{table*}

For this demonstration we choose to use the equatorial field of the XMM-XXL survey \citep{Pierre2016, Liu2016}, because of the rich set of multi-epoch observations over an 18-year baseline. They include some of the largest extragalactic survey programmes carried out by \textit{XMM-Newton}, such as the XMM Large Scale Structure survey \citep[XMM-LSS;][]{Pierre2007}, the Subaru/{\it XMM-Newton} Deep Survey \citep[SXDS;][]{Ueda2008}, the XMM-XXL survey \citep{Pierre2016} and the XMM-SERVS \citep[Spitzer Extragalactic Representative Volume Survey;][]{Chen2018}. Additionally, this field benefits from deep photometric observations and extensive follow-up spectroscopic programmes \citep[e.g. see][]{Georgakakis2017xxl}, some of which have targeted specifically X-ray sources \citep[e.g.][]{Stalin2010, Menzel2016}. These ancillary data facilitate the multiwavelength characterisation of interesting variable sources. The input positions to the RapidXMM server are selected from the 4XMM-DR9 catalogue after applying spatial filters to isolate sources in the area of the XMM-XXL field, a detection likelihood cut to minimise spurious sources, an X-ray extension flag to exclude extended emission and an off-axis angle limit to avoid artifacts close to the edge of the field-of-view. These criteria are listed in Table \ref{tab:xxl-sel}. Only EPIC-PN detections are considered because of the higher sensitivity of this camera. We choose the 0.2--2\,keV band to compare the 4XMM-DR9 fluxes with the RapidXMM upper limits because of the higher effective area of \textit{XMM-Newton} at these energies, which translates to higher sensitivity. The 4XMM-DR9 photometry in this band is estimated as the sum of the count rates in the 4XMM-DR9 bands 1 (0.2--0.5\,keV), 2 (0.5--1\,keV) and 3 (1--2\,keV). 

Figure \ref{fig:xxl_ff} presents the comparison between the 4XMM-DR9 catalogued count-rates against those determined by querying the RapidXMM database. All the datapoints on this figure are  4XMM-DR9 sources in the XMM-XXL field that fulfil the selection criteria of Table~\ref{tab:xxl-sel}. The RapidXMM photometry is determined at the position of each 4XMM-DR9 sources and separately for all the multi-epoch Obs.IDs that these sources lie on. For Obs.IDs on which a given 4XMM-DR9 source is not formally detected by the PPS, the RapidXMM photometry plotted in Fig.~\ref{fig:xxl_ff} corresponds to the $3\sigma$ upper limit (red triangles). For the same Obs.ID as the one on which a given 4XMM-DR9 source is detected the RapidXMM count-rate is estimated via Eq.~\ref{eq:CR_RAPID} (blue dots). The latter set of points traces the one-to-one count-rate relation and shows the good agreement between the 4XMM-DR9 and RapidXMM photometry for the same Obs.ID. This point has already been demonstrated in Fig.~\ref{fig:4xmm_ff_pn} for sources selected in the full 4XMM-DR9 catalogue. The bulk of the upper-limits in Fig.~\ref{fig:xxl_ff} lie above or close to the one-to-one relation. This is the case of multi-epoch  observations that overlap with the position of a given 4XMM-DR9 source and are either shallower or have a higher background level than the Obs.ID on which the source in question is actually detected. There is nevertheless a small fraction of upper limits that scatter below the one-to-one relation. These are potentially interesting since they may include strongly variable sources. Spurious 4XMM-DR9 detections however, which inevitably exist in any source catalogue, will also crop up in the same region of the parameter space. Visual inspection is therefore necessary to identify bona-fide vanishing sources. For this demonstration we choose to eyeball the sources with the larger deviation from the  one-to-one relation by applying an arbitrary threshold of $>4$ to the ratio between the 4XMM-DR9 count rate and the $3\sigma$ upper limit stored in the RapidXMM database. These are the sources (total of 23) that lie below the dot-dashed line in Fig.~\ref{fig:xxl_ff}. Many of them are spurious 4XMM-DR9 detections associated with hot-pixels or high-background regions.  Table~\ref{tab:xxl-sources} presents X-ray information on the bona-fide vanishing sources with flux ratio $>4$ following the visual inspection. The X-ray thumbnails for these sources are shown in Fig.~\ref{fig:xxl-thumb}. This figure includes two X-ray images for each 4XMM-DR9 source, one from the Obs.ID on which it is detected, and the second from the Obs.ID on which it disappears. Optical spectroscopy from the various SDSS spectroscopic surveys in the XMM-XXL area \citep[see][]{Menzel2016} is available for two of the sources listed in Table \ref{tab:variability}, 4XMM ID 207804507010009, 201126801010013. Both of them are AGN at moderate redshifts, $z=0.601$ and 0.492 respectively. It is also interesting to note that two of the sources in Table \ref{tab:variability} vary by more than 1\,dex in flux over a period of nearly 14 years.

\subsection{Stacking}
\label{sec:stacking}
Stacking techniques are frequently used in X-ray astronomy to explore the mean X-ray properties of an astrophysical population of sources, whose individual members are too X-ray faint to be individually detected.  Stacking is the accumulation of X-ray photons at the positions of sources that lie below the detection threshold of a given dataset. Given large enough numbers of input sources, this approach may yield a statistically significant, yet intrinsically weak, astrophysical signal that places constraints on the ensemble X-ray properties of the population, e.g. mean flux.  This principle has been applied to both X-ray images that accumulate X-ray photons in fixed energy bands \citep[e.g.][]{Georgakakis03, Hickox07a, Hickox07b, Georgakakis2008_stack, Chen13} and X-ray spectra \citep{Streblyanska05, Corral08, Falocco14}. In this paper we focus on the former. This is equivalent to performing aperture photometry at the sources’ positions. Image sections (e.g. apertures) in the vicinity of the sources of interest are extracted and coadded to increase the signal-to-noise ratio (SNR). The resulting stacked image can then be analysed to yield information on the average flux or X-ray hardness of the population. It is emphasised that the X-ray stacking described above is different from the coaddition of overlapping {\it XMM-Newton} observations to improve the statistics of individual sources and detect fainter ones \citep[e.g. XMM Stacked catalogue,][]{Traulsen2019}. Our application coadds the X-ray photons at input positions spread across multiple and typically non-overlapping observations.

The RapidXMM database is well suited for stacking analysis because (i) all data needed for X-ray aperture photometry (counts around the positions of interest, background estimations and exposure times) are readily available, and (ii) the data is organised in a regular grid (the HEALPix tessellation) that facilitates the extraction of counts at different sky positions and their coaddition via the translation into a new reference frame with a common centre. We demonstrate these features of the RapidXMM database using a sample of optically selected QSOs identified in the SDSS \citep{Lyke2020} that lie within the footprint of the XMM-XXL. Only the EPIC-PN camera and Pointed observations are used in the analysis. 

\begin{figure}
\begin{center}
\includegraphics[angle=0,width=.45\textwidth]{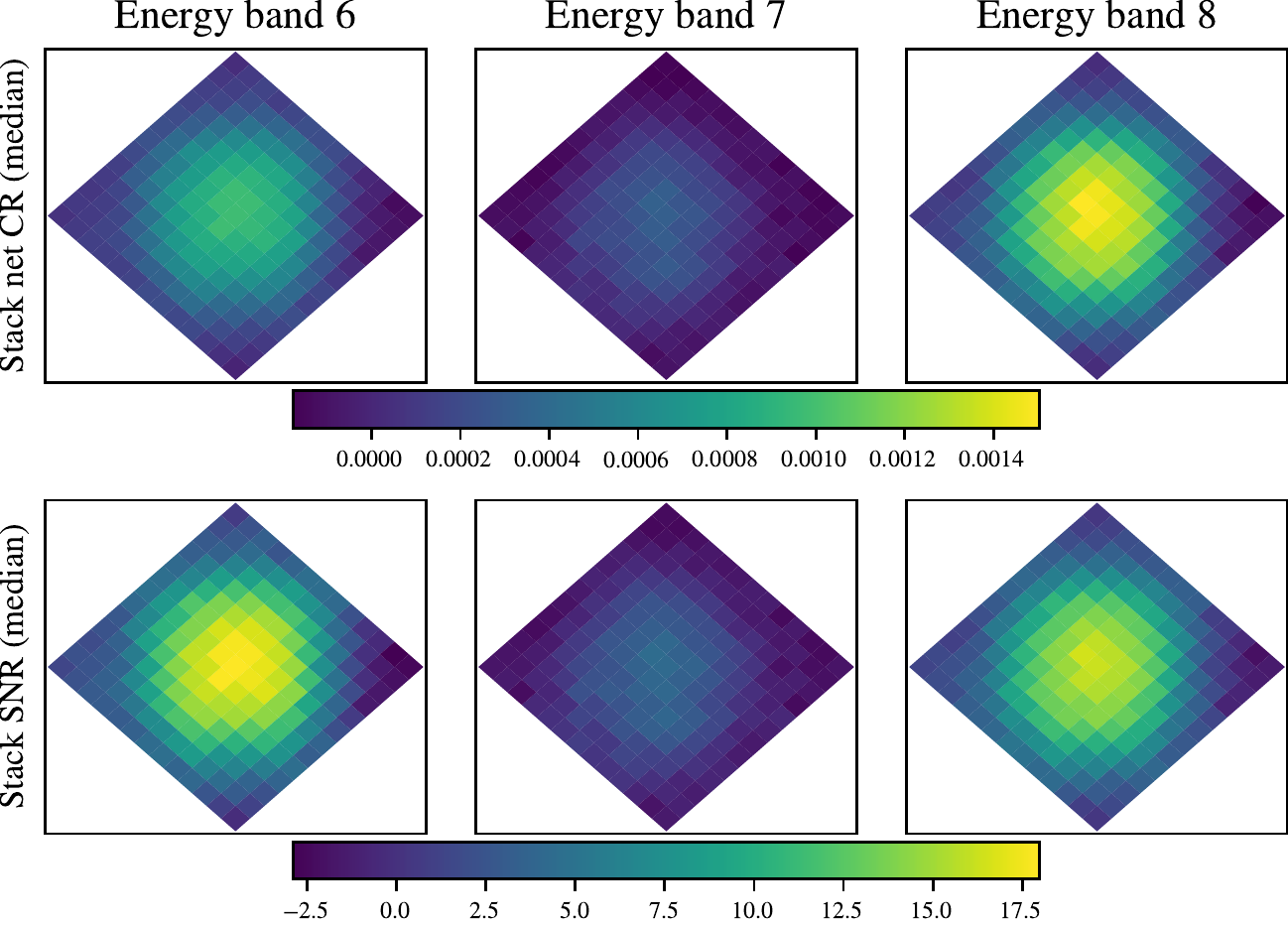}
\end{center}
\caption{Demonstration of stacking analysis using the RapidXMM database. The images on the top show the coadded net count-rate at the positions of 255 non-X-ray detected SDSS QSOs selected in the XMM-XXL field. Each image corresponds to a HEALPix grid with size of $13\times13$ cells. The cells have a linear size of about 3\,arcsec. Each panel corresponds to one of the RapidXMM energy bands, 0.2--2keV (left), 2--12keV (middle) and 0.2--12keV (right). The bottom set of images shows the corresponding SNR of the stacked signal in each of the above energy bands.  }\label{fig:stk_qso_xxl}
\end{figure}

\begin{figure}
\begin{center}
\includegraphics[angle=0,width=.45\textwidth]{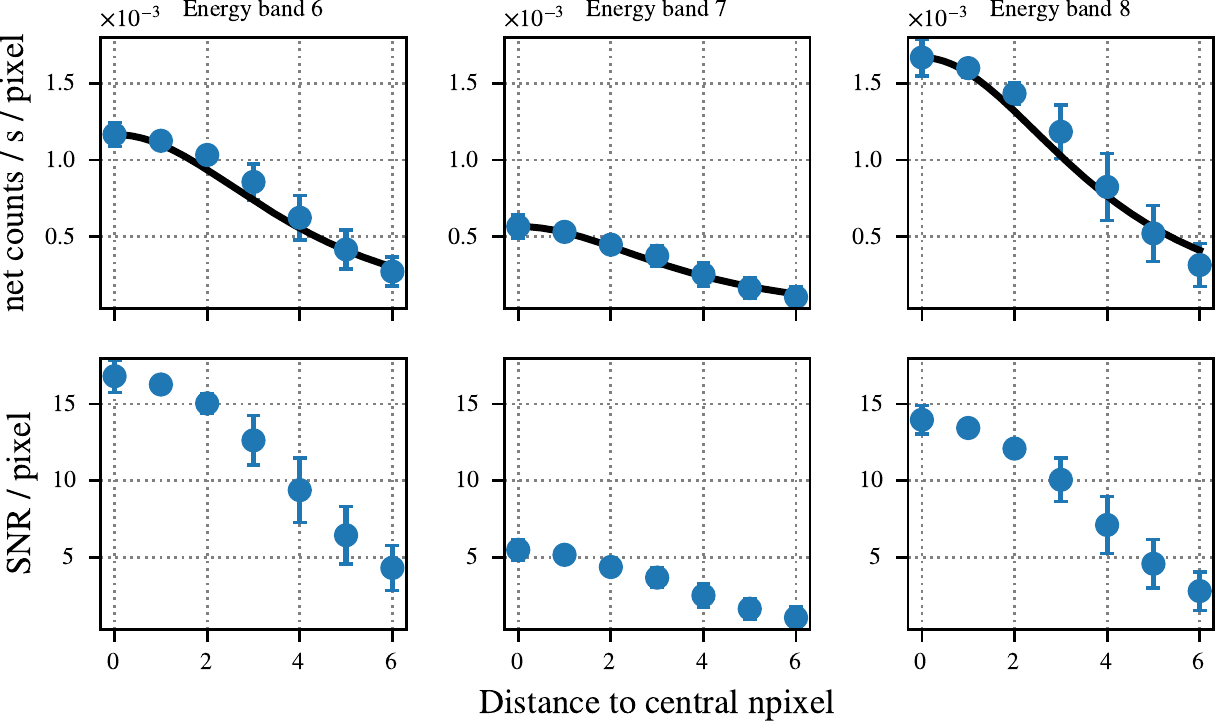}
\end{center}
\caption{Radial profiles of the net-count rate (top panels) and SNR (bottom panels) derived from the images shown in Fig.~\ref{fig:stk_qso_xxl}. The distance on the horizontal axes corresponds to the number of HEALPix cells to the central position.  The count rate in the top set of panels is in units of $10^{-3}\,\rm sec^{-1}\, pixel^{-1}$. The error bars in both panels are the standard deviations estimated from all the cells at a given radial bin. It is only the first data point at the radial distance of zero that is relevant for the stacking. The rest of the radial distribution data points serve visualisation purposes. The black, solid line in the upper set of panels shows the 1D King profile model for the \textit{XMM-Newton} on-axis PSF broadened by a 15\,arcsec radius kernel at 0.8, 8.0, and 4.25 keV for energy bands 6, 7, and 8, respectively.}\label{fig:stk_qso_xxl_radial}
\end{figure}

There are a total of 3329 QSOs in \cite{Lyke2020} catalogue that are within the sky region observed by XMM-XXL. These are split into two samples. The first contains 2917 QSOs that have a 4XMM-DR9 counterpart within 5\,arcsec of the optical position. This is defined as the X-ray detected subsample. The second group consists of 255 objects in regions observed by \textit{XMM-Newton} but far from any 4XMM source ($> 1$\,arcmin) to avoid contamination issues. The latter class of sources is the focus of the stacking analysis, which proceeds as follows. For each optical QSO we find the unique HEALpix cell number {\sc npixel} ({\sc nside}=16 for Pointed observations) that overlaps with the source's position.

These HEALPix cells contain all necessary information to assess whether a stacked signal is detected at the "combined" sky positions of the QSOs, since the counts and associated products for each cell correspond to a circular area of 15 arcsec radius at the nominal coordinates of the cell. Nevertheless, in order to facilitate the visualisation of the stacking results, we use a grid of the k-nearest neighbouring HEALPix cells, with k=6. These are the indices of the topological immediate neighbours of the central pixel out to a distance of 6 cells. The result is a $13\times13$ grid of HEALPix cells around the optical QSO position, corresponding to a region of $\sim43\times43$ arcsec. This is large enough to include the \textit{XMM-Newton}'s PSF, which has a half energy width of $\sim15$ arcsec. A graphical representation of this grid is shown in Figure \ref{fig:stk_qso_xxl}. The RapidXMM database is then queried at the positions of all the cells in the $13\times13$ grid to retrieve the corresponding counts, background levels and mean exposure times. 

\begin{figure}
\begin{center}
\includegraphics[angle=0,width=.45\textwidth]{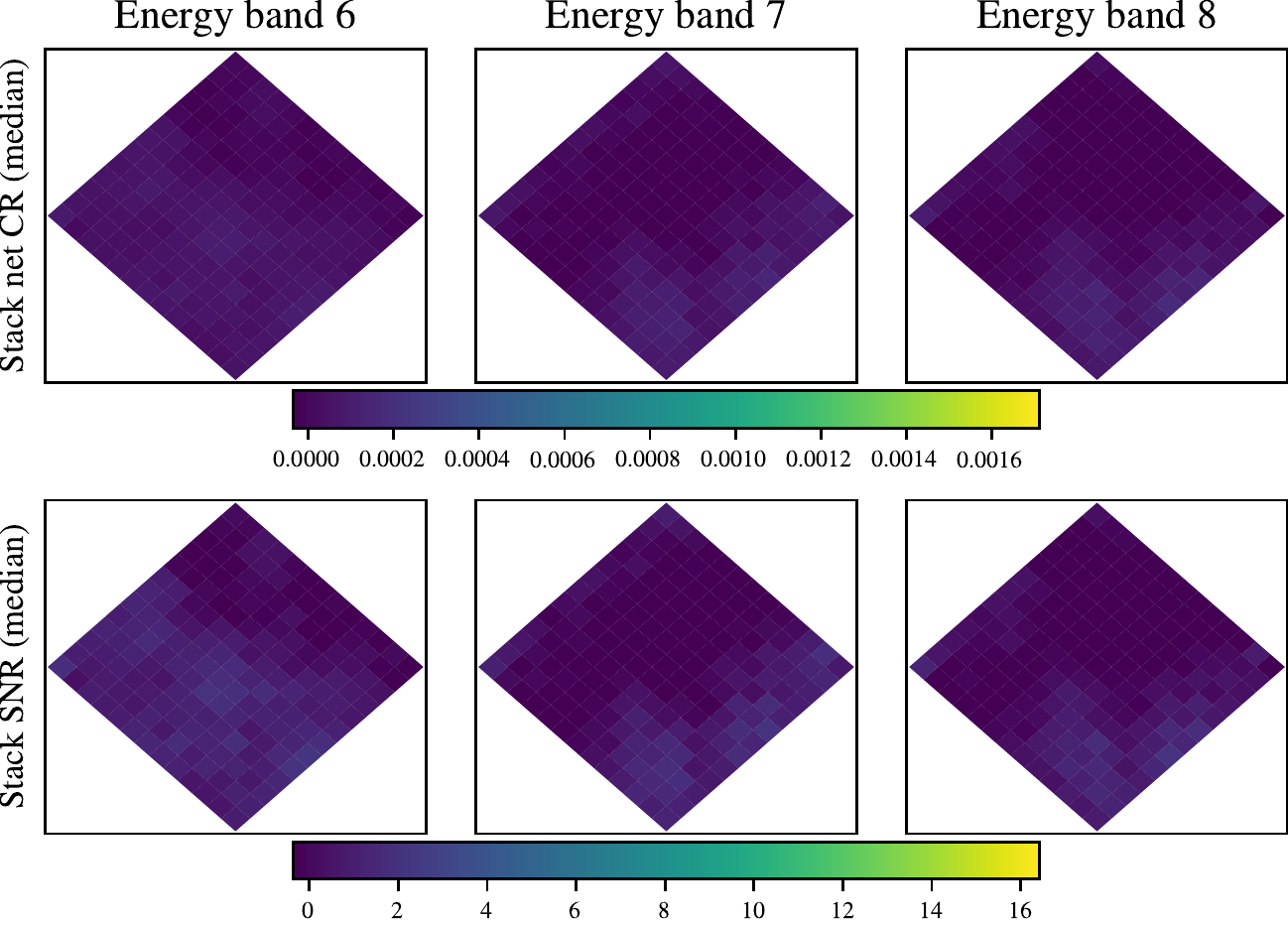}
\end{center}
\caption{Same as in Fig.~\ref{fig:stk_qso_xxl}, but using random positions in the XXL field, far ($>1$\,arcmin) from 4XMM sources or SDSS QSO.}\label{fig:stk_rnd_xxl}
\end{figure}

For the estimation of the local background level for a given QSO position we draw 100 random  HEALPix cells that lie between 60 and 120\,arcsec off the central position and on the same \textit{XMM-Newton} observation as the QSO in question. The RapidXMM database is queried at these random positions to retrieve the corresponding number of counts. The background level is the average number of counts at the random positions after excluding any flagged cells. The background values obtained from this approach are Poisson variates. This allows the robust determination of the background-level uncertainty, which is important for estimating the signal-to-noise ratio of the stacked signal.

Counts, background levels and mean exposure times are organised in cubes with shapes $(13, 13, n)$, where $n$ is the total number of sources in the stack. The optical QSOs may overlap with multiple \textit{XMM-Newton} observations and therefore $n$ is larger than the original size of the sample selected for stacking.  Each of the  RapidXMM energy bands, 0.2--2, 2--12 and 0.2--12~keV, are assigned a different set of cubes. These arrays are then used to estimate averages by integrating along the 3rd ($n$) dimension. A bootstrapping methodology is adopted to minimise the impact of outliers on the final signal. We randomly draw source samples of size $m$ with replacement. For this application $m$ is set to $n$, the total number of sources in the stacking sample. We then coadd the counts, background levels and exposure times of each randomly drawn sample. These coadded values are used to calculate the count-rate via Eq.~\ref{eq:CR_RAPID}. This process is repeated 1000 times to estimate the median count-rate and the median absolute deviation (MAD;  with respect to the median of the count-rate distribution). The MAD serves as a measure robust to outliers of the statistical uncertainty of the stacked signal. The SNR of the stack is estimated as
\begin{equation}\label{eq:SNR}
\begin{split}
SNR & = (S - B) / \sqrt{\sigma_S^2 + \sigma_B^2}, \\
\sigma_S^2 & = S, \\
\sigma_B^2 & = \sum_{i=1}^{N_{QSO}} \delta B^2_{i} =  \sum_{i=1}^{N_{QSO}} \frac{1}{NB_i}\,\sum_{j=1}^{NB_{i}} B_{i,j},
\end{split}
\end{equation}
where $S$ and $B$ represent, respectively, the sums of the counts and background levels at the positions of stacked QSOs. The parameters $\sigma_S$ and $\sigma_B$ are the uncertainties of $S$ and $B$, respectively, assuming Poisson statistics. In this case, $\sigma_B$ is the sum of the local background variances, $\delta B^2_i$, for the $\rm N_{QSO}$ QSOs used in the stacking analysis. Under the assumption of Poisson statistics $\delta B^2_i$ is estimated as the average of the counts extracted from the $NB_{i}$ background cells in the vicinity of the QSO position $i$.

\begin{figure}
\begin{center}
\includegraphics[angle=0,width=.45\textwidth]{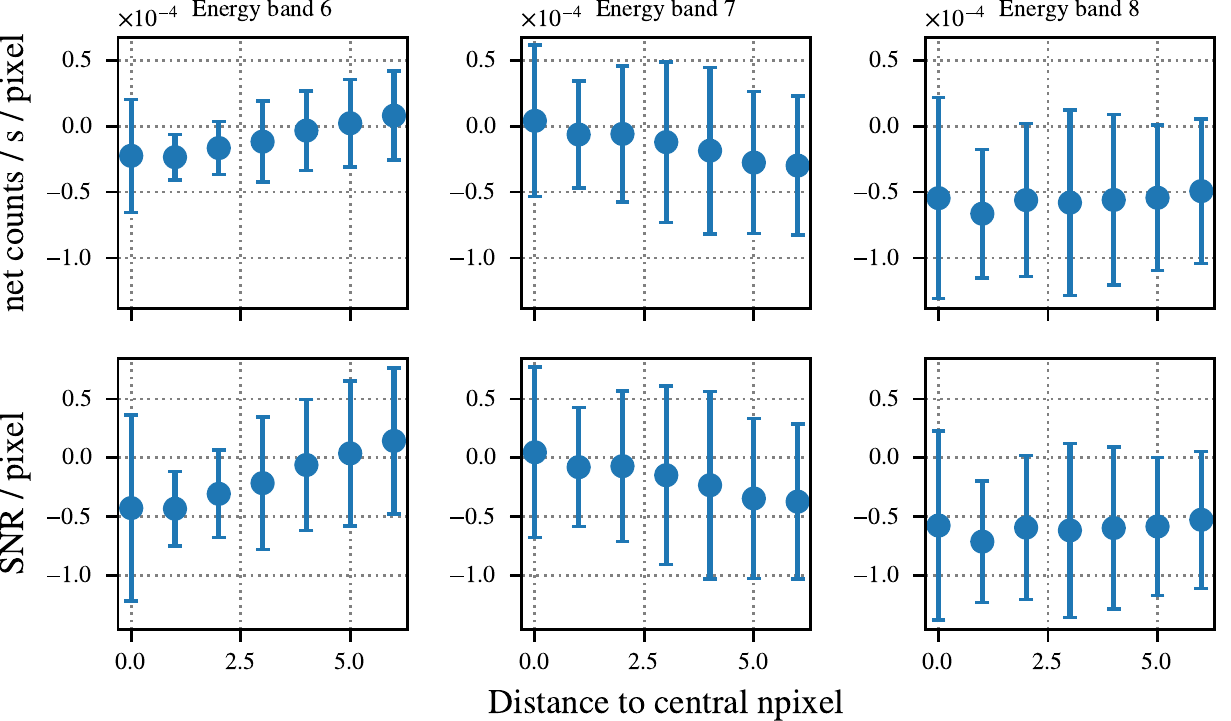}
\end{center}
\caption{Radial profiles of the net-count rate (top panels) and SNR (bottom panels) derived from the images shown in Fig.~\ref{fig:stk_rnd_xxl}. The distance on the horizontal axes corresponds to the number of HEALPix cells to the central position. The count rate in the top set of panels is in units of $10^{-4}\,\rm sec^{-1} \, pixel^{-1}$, i.e. 1\,dex lower compared to the corresponding axis of Figure \ref{fig:stk_qso_xxl_radial}. The error bars are the standard deviations estimated from all the cells at a given radial bin. Only the first data point at a radial distance of zero is relevant for the stacking. The rest of the radial distribution data points serve visualisation purposes.}\label{fig:stk_rnd_xxl_radial}
\end{figure}

Figure~\ref{fig:stk_qso_xxl} shows the results of this method. Each panel of this figure shows the $13\times13$ array of HEALPix cells in each of the 3 energy bands of RapidXMM. The cells are about 3\,arcsec on the side. In the top set of panels the value of each cell corresponds to the median count-rate at that position determined via the bootstrapping approach described above. The panels at the bottom row are the $13\times13$ HEALPix grid of the SNR of each cell. It is emphasised that the values of the cells in Fig.~\ref{fig:stk_qso_xxl} are not independent, since they correspond to the integrated counts at that position within an aperture of 15\,arcsec in radius. Nevertheless, the adopted representation using a $13\times13$ HEALPix grid helps visualise the results.

This figure shows that the stacking signal is maximum close to the centre of the grid, i.e. at the QSO positions, and then gradually drops toward the edges. This trend is more evident in the soft (0.2--2\,keV) and total (0.2--12\,keV) bands and less pronounced in the 2--12\,keV energy range. This dependence on energy of the amplitude of the stacked signal has to do with sensitivity of the EPIC-PN camera of \textit{XMM-Newton} as well as the soft X-ray spectral shape of the optical QSO population that is being stacked. The significance of the stacked signal is further explored in Fig.~\ref{fig:stk_qso_xxl_radial}, which plots the radial profile of the SNR. 

The first radial bin in this figure (plotted at zero distance) corresponds to the HEALPix cell centred at the QSO positions. The data points at radial distances greater than zero are estimated by averaging the values of the HEALPix cells within annuli of increasing radius. It is emphasised that it is the first radial bin plotted at zero distance that contains all the information on the significance of the stacked signal, i.e. signal-to-noise ratio within an aperture of 15\,arcsec centred at the QSO positions. This data point shows that there is a strong signal in the soft and total bands, $\mathrm{SNR}\gtrsim 10$. The signal is less significant in the hard band with $\mathrm{SNR}\sim 5$. The radial bins at distances greater than zero do not provide additional information on the significance. Instead they show that the stacked signal follows well the (convolved) PSF profile. This is demonstrated by overplotting in  Fig.~\ref{fig:stk_qso_xxl} the 1-D model of the on-axis EPIC-PN PSF provided in the \textit{XMM-Newton}'s calibration products.\footnote{\url{https://xmmweb.esac.esa.int/docs/documents/CAL-TN-0029-1-0.pdf}}

For comparison, Figs.~\ref{fig:stk_rnd_xxl} and \ref{fig:stk_rnd_xxl_radial} show the stacking results using as input random positions selected in regions of the XXL far ($> 30$\,arcsec) from X-ray sources or optical QSO. The size of this random sample is equal to the number of the XMM-XXL optical QSOs used to derive the stacking results shown in  Figure~\ref{fig:stk_qso_xxl}. In the case of random positions the stacked signal is consistent with zero within the $1\,\sigma$ uncertainty. This is shown by the SNR of the first radial bin in Figure \ref{fig:stk_rnd_xxl_radial}, which contains all the information on the stacking results. The remaining data points in that figure at larger radial bins further demonstrate the lack of central concentration of the signal, unlike the radial distribution in Figure~\ref{fig:stk_qso_xxl_radial}. The radial profiles of Figure \ref{fig:stk_rnd_xxl_radial} are dominated by the spatially non-uniform background of \textit{XMM-Newton} because of e.g. vignetting, PSF wings, gaps etc.

\begin{figure}
\begin{center}
\includegraphics[angle=0,width=\columnwidth]{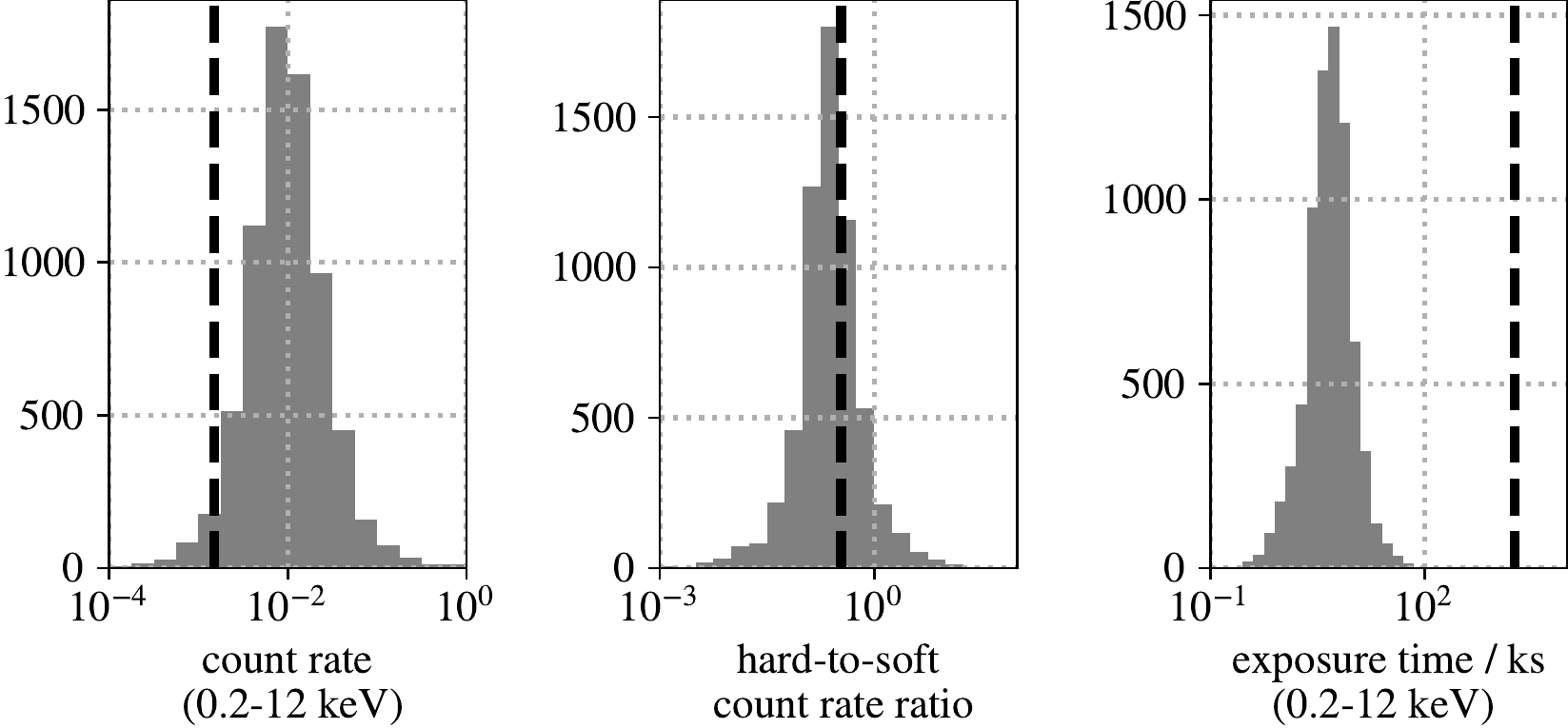}
\end{center}
\caption{Distribution of count rates in the total energy band (left), the hard-to-soft count rate ratios (middle) and exposure times (right) for the SDSS QSO with a 4XMM counterpart. The black dashed lines show the corresponding median value of the stacking of QSO with no X-ray counterpart.}
\label{fig:stk_qso_xxl_crdist}
\end{figure}

Figure~\ref{fig:stk_qso_xxl_crdist} shows the count-rate distribution in the 0.2--12~keV band (left), the hard-to-soft count ratio distribution (middle) and the exposure time distribution (right) for QSO detected in X-rays, and the corresponding median value calculated for the X-ray undetected population using our stacking analysis. As expected, it shows that the median count-rate of the X-ray undetected QSOs are lower, [$(1.7\pm 0.2)\times10^{-3} \mathrm{counts \, sec^{-1}}$] and its median exposure time higher (1800 ks) than the mode of the X-ray detected population. On the other hand, the median hard-to-soft count ratio is similar in both populations, suggesting that both X-ray detected and undetected SDSS QSO have similar spectral shapes.

\section{Conclusions}
We present the development of the infrastructure for a new  \textit{XMM-Newton}  X-ray upper-limit and X-ray photometry server that is available through the XSA.  We describe the pipeline that analyses \textit{XMM-Newton}  observations to produce  count-rate upper limits and aperture photometry in three energy bands (0.2--2~keV, 2--12~keV and 0.2--12~keV) for both the Pointed and Slew Survey data. The design of the database that stores these products is also discussed. The approach adopted for optimising the database for queries by sky position is based on the HEALPix tessellation of the celestial sphere. The X-ray upper limits and photometry products are  projected onto HEALPix grids with spatial resolutions of about 3\,arcsec in the case of Pointed observations and 6\,arcsec for the Slew Survey. They are then stored into database tables that are indexed by the unique integer number of the HEALPix cells. Queries by sky-position are therefore reduced to an integer matching exercise, thereby enabling fast response times for large numbers of input positions. 

We demonstrate science applications enabled by this infrastructure. We show how queries to the database at the positions of 4XMM sources can identify highly variable X-ray populations. It is further demonstrated how to perform X-ray stacking analysis of X-ray faint populations.  We identify a total of 255 SDSS QSOs within the XMM-XXL footprint without counterparts in the 4XMM catalogue. The photons in the database at the positions of these sources are stacked to yield a statistically significant signal.

\section*{Acknowledgements}
We thank the referee for their careful reading of the paper and their constructive comments. The authors acknowledge funding from the European Space Agency through the contract 4000127837/19/ES/JD.
\\
We acknowledge support by the European Innovative Training Network (ITN) "BiD4BEST"\footnote{\url{www.bid4best.org}} (Big Data applications for Black hole Evolution Studies) that received funding by the European Union’s Horizon 2020 research and innovation programme under the Marie Skłodowska-Curie grant agreement No 860744
\\
Based on observations obtained with \textit{XMM-Newton}, an ESA science mission with instruments and contributions directly funded by ESA Member States and NASA.
\\
This research has made use of data obtained from the 4XMM \textit{XMM-Newton} serendipitous source catalogue compiled by the 10 institutes of the \textit{XMM-Newton} Survey Science Centre selected by ESA. This research has made use of data obtained from XMMSL2, the Second \textit{XMM-Newton} Slew Survey Catalogue, produced by members of the XMM SOC, the EPIC consortium, and using work carried out in the context of the EXTraS project ("Exploring the X-ray Transient and variable Sky", funded from the EU's Seventh Framework Programme under grant agreement no. 607452). 
\\
This research made use of Astropy,\footnote{\url{www.astropy.org}} a community-developed core Python package for Astronomy \citep{astropy2018}.

\section*{Data Availability}
\label{sec:xsa}

The RapidXMM database is currently accessible through the XSA\footnote{\url{https://nxsa.esac.esa.int/nxsa-web/}} browser interface. In addition to the RapidXMM products presented in Appendix~\ref{sec:db-fields}, the XSA web interface to the database also returns flux upper limits. These are determined by converting the RapidXMM count-rates to fluxes for a power-law X-ray spectrum with slope $\Gamma=1.7$ absorbed by a Galactic column density $N_H=\rm 3 \times 10^{20}\, cm^2$. These conversion factors dependent on both the camera and  the filter used for a given observation and can be found at the XSA website.\footnote{\url{www.cosmos.esa.int/web/xmm-newton/epic-upper-limits}}

The RapidXMM database can also be queried programatically using HTTP GET requests via \url{http://nxsa.esac.esa.int/nxsa-sl/servlet/get-uls}. Only ra/dec queries are allowed through this method. The output is in JSON format \citep{pezoa2016} following the structure presented in Appendix~\ref{sec:db-fields}. A Python module for querying the database, and the code for the stacking analysis presented in Sect.~\ref{sec:stacking} are available in github ({\url{https://github.com/ruizca/rapidxmm}}). The module allows for both positional and npixel queries, returning Astropy tables as outputs. In the future the RapidXMM database will also be accessible via HILIGT.

\bibliographystyle{mnras}
\bibliography{mybib} 

\appendix

\section{RapidXMM Database Schema}
\label{sec:db-fields}
The RapidXMM database consists of two tables with the same field structure, one storing data for the Pointed observations and the other with the Slew Survey observations. Each table contains 25 fields, as follows:

\smallskip

\noindent
{\sc npixel} (big integer\footnote{In the PostgreSQL implementation used for the RapidXMM database, a big integer is an eight-byte integer (int64) with values between $-9223372036854775808$ and $+9223372036854775807$}): HEALPix cell number using the nested numbering scheme. It corresponds to a HEALPix resolution of {\sc order}=16 for the Pointed observations and {\sc order}=15 for the Slew Survey data. 
  
\noindent  
{\sc obsid} (big integer): Unique Obs.ID number that the npixel and corresponding upper limit belongs to. It is emphasised that this is an integer not a string and therefore the leading zero that characterises the Obs.ID numbers of Pointed Observations is removed.

\noindent
{\sc start\_date} (integer\footnote{In the PostgreSQL implementation used for the RapidXMM database, an integer is a four bytes integer with values between $-2147483648$ and $+2147483647$}): the starting date of the observation in Modified Julian Date.

\noindent
{\sc end\_date} (integer): the end date of the observation in Modified Julian Date.

\noindent
{\sc eef} (real):  Encircled Energy Fraction used for the determination of the upper limits (see Table~\ref{tab:params}).

\noindent
{\sc area\_ratio} (real): Ratio between the number of good pixels in the source extraction region and the number of good pixels in the background extraction region.

\noindent
{\sc band6\_exposure} (real): average exposure time (corrected of vignetting) for the energy band 6 (0.2--2~kev).

\noindent
{\sc band6\_src\_counts} (integer): 
image counts for the energy band 6 (0.2--2~keV).

\noindent
{\sc band6\_bck\_counts} (real): background expectation value scaled to the source extraction region for the energy band 6 (0.2--2~keV).

\noindent
{\sc band6\_ul\_sigma1} (real): $1\sigma$ (one sided, 84.13\%) count-rate upper limit
for the energy band 6 (0.2--2~keV).

\noindent
{\sc band6\_ul\_sigma2} (real): $2\sigma$ (one sided, 97.72\%) count-rate upper limit
for the energy band 6 (0.2--2~keV).

\noindent
{\sc band6\_ul\_sigma3} (real): $3\sigma$ (one sided, 99.87\%) count-rate upper limit
for the energy band 6 (0.2--2~keV).

\noindent
{\sc band7\_exposure} (real): average exposure time (corrected of vignetting) for the energy band 7 (2--12~keV).

\noindent
{\sc band7\_src\_counts} (integer): image counts for energy band 7 (2--12~keV).

\noindent
{\sc band7\_bck\_counts} (real): background expectation value  scaled to the source extraction region for the energy band 7 (2--12~keV).

\noindent
{\sc band7\_ul\_sigma1} (real): $1\sigma$ (one sided, 84.13\%) count-rate upper limit
for the energy band 7 (2--12~keV).

\noindent
{\sc band7\_ul\_sigma2} (real): $2\sigma$ (one sided, 97.72\%) count-rate upper limit
for the energy band 7 (2--12~keV).

\noindent
{\sc band7\_ul\_sigma3} (real): $3\sigma$ (one sided, 99.87\%) count-rate upper limit
for the energy band 7 (2--12~keV).

\noindent
{\sc band8\_exposure} (real): average exposure time (corrected of vignetting) for the energy band 8 (0.2-12~keV).

\noindent
{\sc band8\_src\_counts} (integer): image counts for the energy band 8 (0.2--12~keV).

\noindent
{\sc band8\_bck\_counts} (real): background expectation value scaled to the source extraction region for the energy band 8 (0.2--12~keV).

\noindent
{\sc band8\_ul\_sigma1} (real): $1\sigma$ (one sided, 84.13\%) count-rate upper limit
for the energy band 8 (0.2--12~keV).

\noindent
{\sc band8\_ul\_sigma2} (real): $2\sigma$ (one sided, 97.72\%) count-rate upper limit
for the energy band 8 (0.2--12~keV).

\noindent
{\sc band8\_ul\_sigma3} (real): $3\sigma$ (one sided, 99.87\%) count-rate upper limit
for the energy band 8 (0.2--12~keV).

\noindent
{\sc bitfields} (integer): 32 bits field containing the quality flags for the upper limits (Sect.~\ref{sec:quality-flag}) and the information about the detector and filter used in the observation (see Table~\ref{tab:bitfields}).

\begin{table}
\centering
\caption{
Description of the {\sc bitfields} bit field.}
\begin{tabular}{ c l}
\hline 
Bit Numbers & Description  \\
\hline
0  & Set to 1 for EMOS1 data \\
1  & Set to 1 for EMOS2 data \\
2  & Set to 1 for EPN data \\
3  & Set to 1 when using no filter \\
4  & Set to 1 when using Thin1 filter \\
5  & Set to 1 when using Thin2 filter \\
6  & Set to 1 when using Medium filter \\
7  & Set to 1 when using Thick filter \\
8--15  & Quality flags for energy band 6 (see Sect.~\ref{sec:quality-flag}) \\
16--23  & Quality flags for energy band 7 (see Sect.~\ref{sec:quality-flag}) \\
24--31 & Quality flags for energy band 8 (see Sect.~\ref{sec:quality-flag}) \\
\hline
\end{tabular}
\label{tab:bitfields}
\end{table}

\bsp	
\label{lastpage}
\end{document}